\documentclass[12pt]{article}
\pdfoutput=1
\usepackage{amsmath,amssymb,mathrsfs,amsbsy,latexsym,amsfonts,amsthm}
\usepackage{hyperref}
\usepackage[utf8]{inputenc}
\usepackage{authblk}
\usepackage{color}
\usepackage{ascmac}
\usepackage{here}
\usepackage{bm}
\usepackage{comment} 
\usepackage[top=30truemm,bottom=30truemm,left=27truemm,right=27truemm]{geometry}
\usepackage{wrapfig}
\usepackage{graphicx}
\usepackage{caption}
\usepackage{enumerate}
\usepackage[italicdiff]{physics}

\usepackage{slashed}
\usepackage{tensor}

\newcommand{\be}{\begin{equation}}
\newcommand{\ee}{\end{equation}}
\newcommand{\ba}[1]{\begin{align}#1\end{align}}

\newcommand{\bfig}{\begin{figure}[htb!] \centering}
\newcommand{\efig}{\end{figure}}

\newcommand{\non}{\nonumber\\[3mm]}

\newcommand{\pa}[1]{\left( #1 \right)}

\newcommand{\nn}{\nonumber\\}

\newcommand{\pc}{\pi_{(c)}}
\newcommand{\bpc}{\bar{\pi}_{(c)}}
\newcommand{\intx}{\int \!\! d^3 x}

\newcommand{\intk}{\int \!\! \frac{d^3 k}{(2\pi)^3}}
\newcommand{\intkw}{\int \widetilde{d^3k}\,}
\newcommand{\intp}{\int \!\! \frac{d^3 p}{(2\pi)^3(2 E_p)}}

\newcommand{\vx}{\vec{x}}
\newcommand{\vy}{\vec{y}}
\newcommand{\vk}{\vec{k}}
\newcommand{\vkp}{\vec{k'}}
\newcommand{\vp}{\vec{p}}

\newcommand{\kx}{\vec{k}\cdot \vec{x}}

\makeatletter

\@addtoreset{equation}{section}
\makeatother

\begin{document}
\title{IR finite $S$-matrix by gauge invariant dressed states}
\author[1]{
	Hayato~Hirai\thanks{\tt hirai@n.kisarazu.ac.jp }
}
\author[2, 3]{
	Sotaro~Sugishita\thanks{\tt sotaro@post.kek.jp}
	\vspace{5mm}
}
\affil[1]{\it\normalsize Natural Science Education, National Institute of Technology, Kisarazu College, 2-11-1 Kiyomidai-Higashi, Kisarazu,
Chiba 292-0041, Japan}
\affil[2]{\it\normalsize Department of Physics and Astronomy, University of Kentucky,\protect \\Lexington, KY 40506, USA}
\affil[3]{\it\normalsize Theory Center, 
High Energy Accelerator Research Organization (KEK),
\protect \\
Tsukuba, Ibaraki 305-0801, Japan}
\setcounter{Maxaffil}{0}

\date{ }

\maketitle

\begin{abstract}
Dressed states were proposed to define the infrared (IR) finite $S$-matrix in QED or gravity. 
We show that the original Kulish-Faddeev dressed states are not enough to cure the IR divergences. 
To illustrate this problem, we consider QED with background currents (Wilson lines). 
This theory is exactly solvable but shares the same IR problems as the full QED. 
We show that naive asymptotic states lead to IR divergences in the $S$-matrix and are also inconsistent with the asymptotic symmetry, even if we add the original Kulish-Faddeev dressing operators. 
We then propose new dressed states which are consistent with the asymptotic symmetry.  
We show that the $S$-matrix for the dressed states is IR finite. 
We finally conclude that appropriate dressed asymptotic states define the IR finite $S$-matrix in the full QED.        
\end{abstract}

\newpage
\setcounter{tocdepth}{2}
\tableofcontents

\newpage

\section{Introduction}

$S$-matrix is the central object for scattering physics in quantum field theories.
However, the conventional $S$-matrix is not well-defined for some theories involving massless particles because of  \textit{infrared (IR) divergences}.
A famous example of such theories is \textit{quantum electrodynamics (QED)} in 4-dimensional Minkowski spacetime.
When we compute the $S$-matrix perturbatively, the contributions from low-energy virtual photons cause divergences of loop diagrams.
Because the infrared divergences cannot be eliminated by any renormalization procedure, the $S$-matrix is  not a well-defined object at any order of perturbation (except for the tree level). 
The resummation of all orders of the IR divergent terms gives exponentially suppressed factors to $S$-matrix elements for any nontrivial processes, and  makes the $S$-matrix trivial \cite{Yennie:1961ad, Weinberg:1965nx, Weinberg:1995mt, Carney:2017jut}. 

The traditional prescription for avoiding the infrared problem is to calculate the \textit{inclusive} cross-sections \cite{Bloch:1937pw}.
In this prescription, we compute the sum of the cross-sections for all processes including possible emissions of real soft bosons (photons, gravitons)  that are physically indistinguishable.
As is well known, the inclusive cross-sections are IR finite. 
The IR divergences caused by virtual soft bosons are canceled out by 
those arising from emissions of real soft bosons.
However, the $S$-matrix itself remains ill-defined in this prescription.

 An alternative approach is to directly construct the well-defined $S$-matrix without IR divergences by using more appropriate asymptotic states instead of Fock states \cite{Chung:1965zza, Greco:1967zza, Kibble:1968sfb, Kibble:1969ip, Kibble:1969ep, Kibble:1969kd, Kulish:1970ut, Greco:1978te}.
In conventional computations of $S$-matrix, it is usually assumed that the asymptotic states for charged particles obey free dynamics although  massless bosons mediate infinitely long-range interactions. 
The key idea of the alternative approach is to use appropriate asymptotic states incorporating the effect of long-range interactions to compute the $S$-matrix.

A candidate for such asymptotic states was first proposed in QED by Chung \cite{Chung:1965zza}.
The states are dressed by an infinite number of coherent soft photons, which are sometimes referred to as \textit{dress} or \textit{cloud} of soft photons.
It was shown that the naive $S$-matrix for the dressed states is infrared finite to all orders perturbatively \cite{Chung:1965zza}. 
Kulish and Faddeev derived other similar dressed states by solving the asymptotic dynamics of QED \cite{Kulish:1970ut}.
Analogous dressed states for perturbative gravity were also obtained by \cite{Ware:2013zja}.
It is also known that the Kulish-Faddeev (KF) dressed states can be obtained by solving the gauge invariant (BRST) condition in asymptotic regions \cite{Hirai:2019gio}.\footnote{In \cite{Kulish:1970ut}, the dressed states derived by solving the infrared dynamics are modified by introducing  an artificial null vector $c_{\mu}$ in order to resolve the problem that the dressed states do not satisfy the conventional Gupta-Bleuler condition  ($k^{\mu}a_{\mu}(k)\ket{\psi}=0$).
 In \cite{Hirai:2019gio}, it is shown that the condition is not adequate for dressed states and  also shown that the unmodified KF dressed states are gauge invariant without introducing the vector $c_{\mu}$.
 In this paper, KF dressed states mean the unmodified states.
 }

Although the dressed states  were proposed many years ago, they have been recently reinvestigated in the connection to the asymptotic symmetry (see, e.g., \cite{Mirbabayi:2016axw, Gabai:2016kuf, Kapec:2017tkm, Choi:2017bna, Choi:2017ylo, Carney:2018ygh, Neuenfeld:2018fdw, Hirai:2019gio, Gonzo:2019fai, Choi:2019rlz,  Choi:2019sjs, Furugori:2020vdl}).
The asymptotic symmetry in QED is a part of large $U(1)$ gauge transformations.
It is a physical symmetry in the sense that the conservation law of the Noether charge associated with the symmetry leads to a nontrivial constraint.
It was pointed out in \cite{Kapec:2017tkm} that the vanishing of $S$-matrix elements in the conventional computations is consistent with the asymptotic symmetry of QED.
Initial and final Fock states used in the conventional computations generally belong to different sectors of the asymptotic symmetry. 
Therefore, the amplitude between them should vanish since otherwise it breaks the conservation law. 
Thus it was argued that dressed states are needed in order to obtain non-vanishing amplitudes \cite{Kapec:2017tkm}.

Let us mention a slight but important difference between Chung's dressed states that are a candidate for the  well-defined $S$-matrix and the KF dressed states that are naturally appeared by solving the dynamics of QED. 
The difference is the existence of the oscillating phase factors taking the form of 
$
\exp\left(\pm i\frac{p\cdot k}{E_{p}}t\right)
$
 in KF dresses (see \eqref{def:Rf}, \eqref{def:Ri}).
Here $p^{\mu}=(E_p,\vec{p})$ is a four momentum of an (anti)electron, $k^{\mu}$ is that of a photon, and $t$ is the time when the initial or final state of the scattering is defined. We will take the limit as $t\rightarrow \pm \infty$ at the end of the calculation.
The phase factors naturally appear in the dress by solving the asymptotic dynamics in QED \cite{Kulish:1970ut} or by requiring the gauge invariance \cite{Hirai:2019gio}.
In \cite{Kulish:1970ut} and the literature, the phase factors are set to one by adopting the approximation:
\begin{align}
\label{eq:phaseapp}
\exp\left(\pm i\frac{p\cdot k}{E_{p}}t\right)\sim 1
\end{align}
for``small" $k$.
If we use this approximation, the KF dressed states become almost identical to Chung's ones. 
However, the approximation \eqref{eq:phaseapp} is valid only for the parameter region
\begin{align}
\left|\frac{p\cdot k}{E_{p}}\right|\ll \left|\frac{1}{t}\right|\,.
\end{align}
When we take the limit $t\rightarrow \pm \infty$, this parameter region disappear. 
Therefore the approximation \eqref{eq:phaseapp} is not justified in this limit.

In this paper, we propose new gauge invariant dressed states which are consistent with the asymptotic symmetry in QED, 
and show that the $S$-matrix for the dressed states is IR finite.
If we remember the derivation of the soft photon theorems \cite{Weinberg:1965nx} or evaluation of IR divergences in the loop diagrams in QED, the leading divergences are caused by soft photons interacting with on-shell charged particles. 
This fact suggests that we can replace the charged current by a background current of point-particles if we focus just on the IR problem.  
Thus we first consider QED with the background current.
In other words,  
we consider the Maxwell theory with background Wilson lines,   
which is exactly solvable but shares the same IR problems as the full QED. 
The dynamics of this theory is actually identical to the ``asymptotic dynamics'' considered by Kulish and Faddeev  \cite{Kulish:1970ut}. 
We see in the model that  usual Fock states lead to IR divergences in the $S$-matrix. 
In addition, the original KF dressed states cannot remove the IR divergences, 
and are also inconsistent with the asymptotic symmetry. 
The consistency with the asymptotic symmetry implies that we should introduce an additional dressing factor.
We confirm that we can use Chung's dressing factor as the additional one. 
After all, we have to use simultaneously both of Chung's dressing factor and KF's one. 
We then show that the $S$-matrix for the new dressed states is IR finite.  
We finally argue that we can apply the new dressed states to the full QED, and conclude that appropriate dressed asymptotic states define the IR finite $S$-matrix.

The rest of this paper is structured as follows.
In section \ref{sec:main}, we resolve the IR problem in QED with a fixed current of point-particles which we call the background current model.
We first explain the model in section \ref{subsec:model} and construct the $S$-matrix in the model in section \ref{subsec:Smatrix}.
In section \ref{subsec:IRdiv}, we see that the same IR divergences as QED appear in the model, and also that the IR divergences cannot be eliminated even if we add the KF dressing operators to initial and finial states.
In section \ref{sec:Gauss}, we obtain physically allowed dressed states by solving the gauge invariant condition.
In section \ref{sec:AS}, we express that the asymptotic symmetry requires an additional dressing factor other than KF's factor. 
We find that the additional factor can be Chung's one. 
In subsection \ref{sec:asyCL}, we present the expressions of charges associated with asymptotic symmetry at asymptotic regions and also comment on their conservation laws.
Section \ref{sec:IRfinite} shows that the $S$-matrix for new dressed states has no IR divergences.
In section \ref{sec:fullQED}, the implications for the full QED are discussed.
Section \ref{sec:conclusion} contains our conclusion  and discussions about future directions.
Appendix \ref{app:BRST} shows the basic results of  BRST quantization of our model.
Appendix \ref{app:as} contains the derivation of charges of asymptotic symmetry in our model.
Appendix \ref{app:falloff} shows that the Li\'enard-Wiechert potential does not contribute to the charge of asymptotic symmetry in the asymptotic limit.
Appendix \ref{app:soft} shows that the current can be approximated by the point-particles current at asymptotic region in QED.

\section{Resolution of IR problem in background current model of QED\label{sec:main}}

In this section, we consider QED with a fixed current of point-particles. 
We call this theory the background current model. 
This model shares  the same IR problem as the full QED, and we see how dressed states resolve it.

\subsection{QED with a fixed current}\label{subsec:model}

\begin{figure}[H]
\centering
\includegraphics[height=6cm]{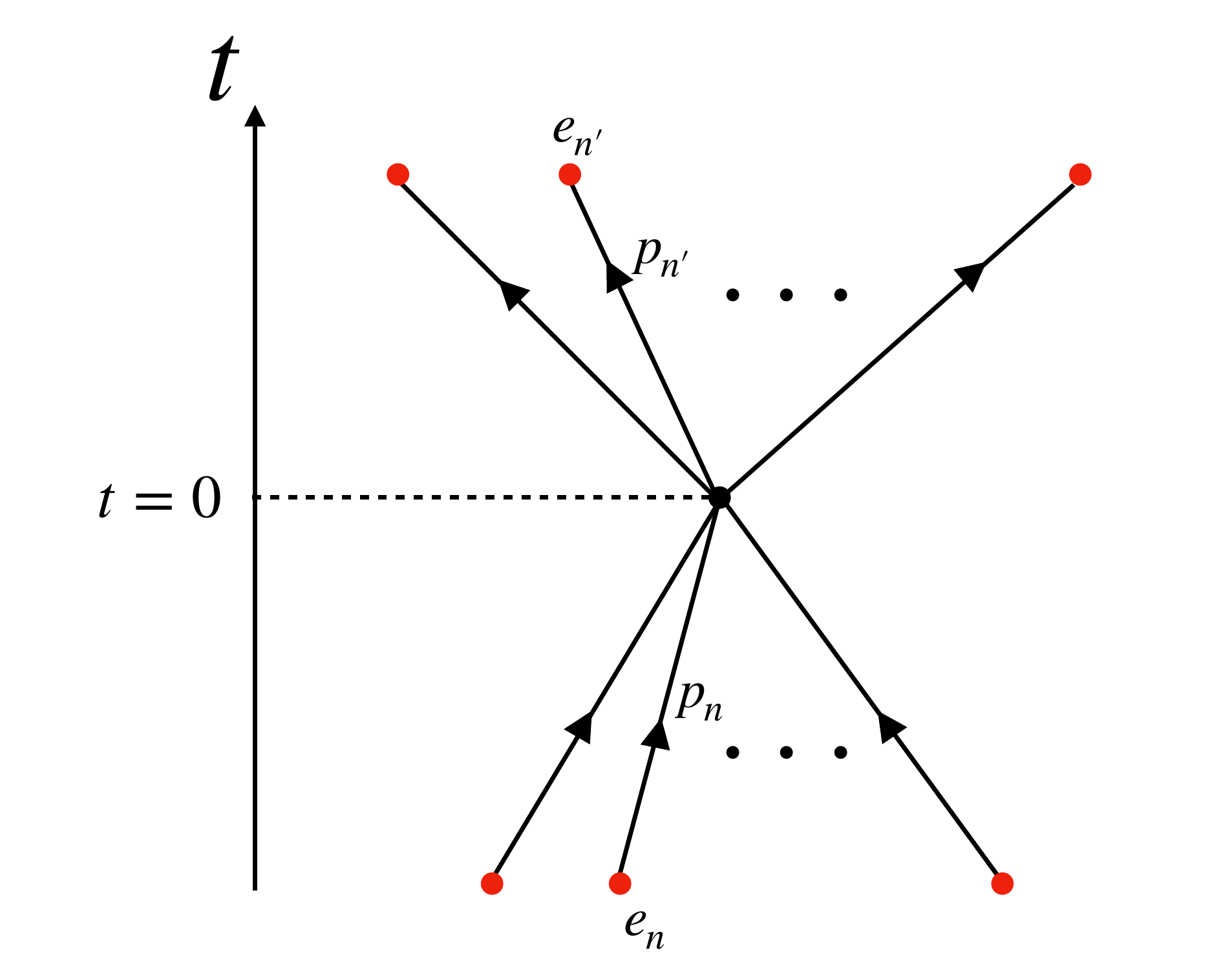}
\caption{The trajectories of charged particles corresponding to \eqref{j_cl}.}
\label{fig:current}
\end{figure}

The IR problem in QED is universal in the sense that the detailed information of scatterings is not important. 
More concretely, the IR divergences of $S$-matrix elements depend only on charges and momenta of the initial and the final states of charged particles.
It is expected that the detailed dynamics of charged particles are not so relevant to understand the IR structure.  
It leads us to  consider QED with a fixed current as
\begin{align}
\mathcal{L}=-\frac{1}{4}F_{\mu\nu}F^{\mu\nu}
+j^\mu_{\mathrm{pp}}A_\mu,
\label{eq:Lag}
\end{align}
where
\begin{align}
\label{j_cl}
j^\mu_{\mathrm{pp}}(x)=\Theta(-t)\sum_{n\in I} \frac{e_n p_n^\mu}{E_{n}}\, \delta^{3}(\vx-\vp_n t/E_{n})
+\Theta(t)\sum_{n \in F} \frac{e_{n} p_{n}^\mu}{E_{n}}\, \delta^{3}(\vx-\vp_{n}t/E_{n}).
\end{align}
This current $j^\mu_{\mathrm{pp}}$ corresponds to uniformly moving charged point-particles. Initially the charged particles have momenta $\{p_n\}_{n\in I}$, then scatter instantaneously at $t=0$, and finally have momenta $\{p_{n}\}_{n\in F}$ (see Fig.~\ref{fig:current}).  
Here, $I$ and $F$ denotes \textit{initial} and \textit{final}.
\vspace{5mm}
We represent the mass of each particle as $m_n$ where $p_n^2=-E_n^2+\vp_n^{~2}=-m_n^2$.
We suppose that the total charge is conserved; $\sum_{n \in I}e_n=\sum_{n \in F}e_{n}$. It ensures the current conservation $\partial_\mu j^\mu_{\mathrm{pp}}=0$.
The trajectories of charged particles in eq.~\eqref{j_cl} mean that we have cusped Wilson lines with the cusp at the origin $x=0$.  It implies that this theory can be a good approximation of the full QED when the charged particles are very massive, and shares the same IR structures with the full QED. 
In fact, we will see that the naive $S$-matrix in the theory \eqref{eq:Lag} has the same IR divergences as in  the full QED for the scatterings of charged particles $\{p_n\}_{n\in I} \to \{p_{n}\}_{n\in F }$ and hard photons. Therefore, this theory is a good toy model to understand the IR problems in QED. Note that the term $j^\mu_{\mathrm{pp}}A_\mu$ in eq.~\eqref{eq:Lag} is actually nothing but the ``asymptotic interaction" considered by Kulish and Faddeev in \cite{Kulish:1970ut}. 

Before proceeding, we briefly comment on the classical counterpart of this theory. 
The theory \eqref{eq:Lag} is the free Maxwell theory with a fixed source. In the  Lorenz gauge, the general solutions of the classical equation of motion, $\Box A^\mu=-j^\mu_{\mathrm{pp}}$ are given by 
\begin{align}
   A^\mu(x)=A^\mu_{in}(x)+\int d^4x' G(x,x')j^\mu_{\mathrm{pp}}(x'),
   \label{eq:cl_sol}
\end{align}
where $A^\mu_{in}$ is an initial configuration satisfying $\Box A^\mu_{in}=0$, and $G(x,x')$ is the retarded Green's function. The second term does not depends on the initial field $A^\mu_{in}$. In this sense, the scattering of electromagnetic waves is trivial in this simplified situation. 
We will see the same result in the quantum case. In the case, the effect of the second term is realized as  a dressing operator (which is a displacement operator creating a coherent state of photons). Except for this dressing effect, the scattering of photons should be trivial. It means that for appropriate dressed asymptotic states the $S$-matrix of the photon sector is trivial in this simple toy model \eqref{eq:Lag}.

\subsection{Dyson's $S$-matrix}\label{subsec:Smatrix}
We consider the time-evolution for the theory \eqref{eq:Lag}. 
In the Feynman gauge, the Hamiltonian in the Schr\"{o}dinger picture\footnote{The superscript $``\mathrm{s}"$ denotes the Schr\"{o}dinger picture. We also use the superscript $``I"$ for the interaction picture.} is given by\footnote{
See section 2 in \cite{Hirai:2019gio} for more details.
}
\begin{align}
    H^\mathrm{s}(t)=H_0+V^\mathrm{s}(t)
\end{align}
where
\begin{align}
 & H_0= \intx \left[\frac12 \Pi^{\mathrm{s}}_\mu \Pi^{\mu \mathrm{s}} +(\partial_i\Pi_0^\mathrm{s})   A^{i\mathrm{s}} +(\partial_i \Pi^{i\mathrm{s}})  A^{0\mathrm{s}} +\frac14 F^\mathrm{s}_{ij} F^{ij \mathrm{s}}\right],\quad\\[3mm]
   &V^\mathrm{s}(t)=-\intx j^\mu_{\mathrm{pp}}(t,\vx)A^{\mathrm{s}}_{\mu }(\vx),
    \label{H0V}
\end{align}
and $\Pi_{\mu}$ are the conjugate momenta of $A^\mu$ satisfying the commutation relation
\begin{align}
    [A_\mu^{\mathrm{s}}(\vx), \Pi_\nu^{\mathrm{s}}(\vx')]=i\eta_{\mu\nu}\delta^3(\vx-\vx').
\end{align}
Note that the Hamiltonian has an explicit time-dependence, even in the Schr\"{o}dinger picture, through the classical background current.  

Introducing the annihilation and creation operators of photons with the commutation relation
\begin{align}
    [a_\mu(\vk),a_\nu^{\dagger}(\vkp)]=(2\omega)\eta_{\mu\nu}(2\pi)^3  \delta^3(\vk-\vkp),
\end{align}
we can write the fields $A^\mu, \Pi^\mu$ as
\begin{align}
\label{a-A}
&A_\mu^\mathrm{s}(\vx)=\intkw \left[a_\mu(\vk)e^{-i\omega t_s +i \kx} +a_\mu^{\dagger}(\vk) e^{i\omega t_s -i \kx}\right],
\\
\label{a-P0}
&\Pi_0^\mathrm{s}(\vx)=-i \intkw \left[k^\mu a_\mu(\vk)e^{-i\omega t_s +i \kx} -k^\mu a_\mu^{\dagger}(\vk) e^{i\omega t_s -i \kx}\right],
\\
\label{a-Pi}
&\Pi_i^\mathrm{s}(\vx)=-i \intkw \left[(k_i a_0(\vk)+\omega a_i(\vk))e^{-i\omega t_s +i \kx} -(k_i a_0^{\dagger}(\vk)+\omega a_i^{\dagger}(\vk))e^{i\omega t_s -i \kx}\right],
\end{align}
where 
\begin{align}
    \intkw:=\int \!\! \frac{d^3 k}{(2\pi)^3(2\omega)}
\end{align}
is the Lorentz invariant measure and $t_s$ is an arbitrary time that defines operators in the Schr\"{o}dinger picture.
The free Hamiltonian is given by (up to the zero point energy)
\begin{align}
    H_0=\intkw \omega\,a_\mu^{\dagger}(\vk)a^{\mu}(\vk),
\end{align}
and the interaction is 
\begin{align}
   V^\mathrm{s}(t)= -\intkw \left[a_{\mu}(\vk)j^\mu_{\mathrm{pp}}(t,-\vk)e^{-i\omega t_s} +a^{\dagger}_{\mu}(\vk)j^\mu_{\mathrm{pp}}(t,\vk) e^{i\omega t_s}\right],
\end{align}
where
\begin{align}
    j^\mu_{\mathrm{pp}}(t,\vk):=\intx \,e^{-i \kx} j^\mu_{\mathrm{pp}}(t,\vx).
\end{align}
The time-evolution operator in the Schr\"{o}dinger picture is given by
\begin{align}
\label{eq:U}
    U(t_1,t_2)=\mathrm{T}\exp\left(-i\int^{t_1}_{t_2} dt H^\mathrm{s}(t)\right),
\end{align}
where $\mathrm{T}$ represents the time-ordering.  
What we want to compute is the amplitude defined as
\begin{align}
\label{eq:defSmatrix}
S_{\alpha,\beta}:=\tensor[_s]{\bra{\alpha(t_f)}}{}U(t_f,t_i)\ket{\beta(t_i)}_s
\end{align}
for scattering states $\ket{\alpha(t_f)}_s$ and $\ket{\beta(t_i)}_s$. 

We also introduce the free time-evolution operator as
\begin{align}
    U_0(t_1,t_2)=e^{-iH_0(t_1-t_2)}.
\end{align}
The time-evolution operator \eqref{eq:U} is the same as that for the ``asymptotic interaction" in  \cite{Kulish:1970ut} and can be computed easily as follows (see also \cite{Hirai:2019gio}).  
Using the M\o ller operator defined by 
\begin{align}
\label{eq:Moller}
    \Omega(t):=U(t_s,t)U_0(t,t_s),
\end{align}
we can write the operator \eqref{eq:U} as 
\begin{align}
   U(t_f,t_i)=U_0(t_f,t_s)\Omega^\dagger(t_f)\Omega(t_i)U_0(t_s,t_i):=U_0(t_f,t_s)S_0(t_f,t_i)U_0(t_s,t_i),
\end{align}
where $S_0(t_f,t_i):=\Omega^\dagger(t_f)\Omega(t_i)$ is  the usual (finite time) $S$-matrix operator in the interaction picture. 
It can be expressed by the Dyson series as
\begin{align}
\label{eq:S0}
    S_0(t_f,t_i)&=\mathrm{T}\exp\left(-i\int^{t_f}_{t_i} dt\, V^I(t)\right),
\end{align}
where $V^I$ is the interaction term in the interaction picture
\begin{align}
      V^I(t)&:= U_0(t,t_s)^{-1} V^\mathrm{s}(t)U_0(t,t_s)=-\intkw \left[a_{\mu}(\vk)j^\mu_{\mathrm{pp}}(t,-\vk)e^{-i\omega t} +a^{\dagger}_{\mu}(\vk)j^\mu_{\mathrm{pp}}(t,\vk) e^{i\omega t}\right].
\end{align}
Since $V^I$ is linear in $a^{\mu}$ or $a^{\mu\dagger}$, the commutator $[V^I(t_1),V^I(t_2)]$ is a c-number function as
\begin{align}
\label{commu_VI}
    [V^I(t_1),V^I(t_2)]=-2i\intkw
   \eta_{\mu\nu} j^{\mu}_{\mathrm{pp}}(t_1,-\vk)j^{ \nu}_{\mathrm{pp}}(t_2,\vk)
    \sin [\omega(t_1-t_2)].
\end{align} 
We can simplify the time-ordering in \eqref{eq:S0} as
\begin{align}
\label{eq:S0-2}
    S_0(t_f,t_i)=e^{i\Phi(t_f,t_i)}e^{-i\int^{t_f}_{t_i} dtV^I(t)},
\end{align}
where
\begin{align}\label{eq:defphase}
    \Phi(t_f,t_i):=\frac{i}{2}\int^{t_f}_{t_i}dt_1 \int^{t_1}_{t_i} dt_2 [V^I(t_1),V^I(t_2)]
\end{align}
which takes a real value because $[V^I(t_1),V^I(t_2)]$  is pure imaginary as \eqref{commu_VI}. 
Thus, $e^{i\Phi(t_f,t_i)}$ is just an oscillating factor with phase $\Phi(t_f,t_i)$.
See \eqref{eq:phase-int} and \eqref{eq:FKphase} in appendix \ref{app:phase} for the concrete expression of the phase $\Phi(t_f,t_i)$.
The expression \eqref{eq:S0-2} is a general formula which holds for any background current. 
Using the explicit form of $j^\mu_{\mathrm{pp}}$ in \eqref{j_cl}, we obtain\footnote{In this paper,  we assume $t_f>0$ and $t_i<0$.} 
\begin{align}
    -i\int^{t_f}_{t_i} dt\, V^I(t)
    =R_{out}(t_f)-R_{out}(0)
    +R_{in}(0)-R_{in}(t_i) 
\end{align}
with 
\begin{align}
\label{def:Rf}
    R_{out}(t)&:=\sum_{n\in F}\intkw \frac{ e_{n} p_{n}^\mu}{p_{n}\cdot k} \left[a_{\mu}(\vk)e^{i\frac{p_{n} \cdot k}{E_n}t}
    -a_\mu^{\dagger}(\vk)
    e^{-i\frac{p_{n} \cdot k}{E_n}t}
    \right],
    \\
    \label{def:Ri}
    R_{in}(t)&:=\sum_{n\in I}\intkw \frac{ e_{n} p_{n}^\mu}{p_{n}\cdot k} \left[a_{\mu}(\vk)e^{i\frac{p_{n} \cdot k}{E_{n}}t}
    -a_\mu^{\dagger}(\vk)
    e^{-i\frac{p_{n} \cdot k}{E_n}t}
    \right].
\end{align}
Note that $R_{out}$ and $R_{in}$ are anti-Hermitian operators, and thus $e^{R_{out}}$ and $e^{R_{in}}$ are unitary operators.

In short,  the time-evolution operator \eqref{eq:U} can be written as
\begin{align}
\label{teo}
   U(t_f,t_i)= 
   U_0(t_f,t_s)e^{i\Phi(t_f,t_i)}
   e^{R_{out}(t_f)+\Delta R-R_{in}(t_i)}
   U_0(t_s,t_i),
\end{align}
where 
\begin{align}
\Delta R:=R_{in}(0)-R_{out}(0)\,.
\end{align}
Using the expression \eqref{teo}, we can write the $S$-matrix in \eqref{eq:defSmatrix} as
\begin{align}
\label{tr-amp}
S_{\alpha,\beta}
   =e^{i\Phi(t_f,t_i)}\,\tensor[_I]{\bra{\alpha(t_f)}}{}
   e^{R_{out}(t_f)+\Delta R-R_{in}(t_i)}\ket{\beta(t_i)}_I\,,
\end{align}
where we have written the states in the interaction picture by using  $\ket{\alpha(t)}_I:=U_0(t_s,t)\ket{\alpha(t)}_s$. 
\subsection{Naive IR divergences}\label{subsec:IRdiv}
Here we see that IR divergences generally appear in the transition amplitude \eqref{tr-amp}  unless we prepare appropriate states $\ket{\alpha}, \ket{\beta}$.

First, the phase $\Phi$ in \eqref{eq:defphase} has terms which diverge in the limit $t_i\rightarrow -\infty, t_f\rightarrow \infty$ as explained in appendix \ref{app:phase}.
The divergent part is given by \eqref{eq:divphase}.
The divergence does not so matter because such a divergent phase can be eliminated by simultaneously redefining the basis states. (See appendix \ref{app:phase} for more concrete discussion.)

Next, we consider the operator $e^{R_{out}(t_f)+\Delta R-R_{in}(t_i)}$ in \eqref{tr-amp}.  
This is  a displacement operator of photons because   $R_{out}$ and $R_{in}$ are anti-Hermitian and  linear in $a_{\mu}$ or $a_{\mu}^{\dagger}$  as \eqref{def:Rf} and \eqref{def:Ri}.
In general a displacement operator
\begin{align}
    D:=\exp\left(
    \intkw \left[C^\mu(\vk)a_\mu(\vk)
    -C^{\mu\ast}(\vk)
    a_\mu^{\dagger}(\vk)
    \right]
    \right)
\end{align}
    can be rewritten into the normal ordering as 
\begin{align}
\label{D}
    D
    &=e^{
    -\frac{1}{2}
    \intkw C^\mu(\vk)
    C_\mu^{\ast}(\vk)}
    e^{-
    \intkw C^{\mu\ast}(\vk)
    a_\mu^{\dagger}(\vk)}
    e^{
    \intkw C^\mu(\vk)a_\mu(\vk)}.
\end{align}    
The first factor may have IR divergences. These divergences are real numbers unlike the divergence in the phase $i\Phi$.

For example,  the transition amplitude between the Fock vacuum for the operator $e^{\Delta R}$ is given by
\begin{align}
\bra{0} e^{\Delta R}\ket{0}
 &= \bra{0}e^{-R_{out}(0)+R_{in}(0)}\ket{0}\non
    &=
    \exp\left(
    -\frac{1}{2}\sum_{n,n'} \eta_n \eta_{n'} e_n e_{n'}
    \intkw \frac{p_n \cdot p_{n'}}{(p_n\cdot k)(p_{n'}\cdot k)}
    \right),
    \label{ampIRdiv}
\end{align}
where $\eta_n=-1$ for incoming particles ($n\in I$) and  $\eta_{n}=1$ for outgoing particles ($n\in F$). 
This integral is IR divergent and in fact  is the same as that appearing in the  $S$-matrix element  for the process $\{p_n\}_{n\in I} \to \{p_{n}\}_{n\in  F}$ in the full QED
(see, \textit{e.g.}, \cite{Weinberg:1995mt}).
As explained in \cite{Weinberg:1995mt}, if we introduce the cutoff in the $k$-integral as $\lambda<|\vk|<\Lambda$, 
\eqref{ampIRdiv} depends on the cutoff as 
\begin{align}
    \exp\left(
    -\frac{1}{2}\sum_{n,n'} \eta_n \eta_{n'} e_n e_{n'}
    \intkw \frac{p_n \cdot p_{n'}}{(p_n\cdot k)(p_{n'}\cdot k)}
    \right)
    \propto   \left(\frac{\lambda}{\Lambda}\right)^A
\end{align}
where $A$ is a positive constant for any nontrivial scattering process \cite{Weinberg:1995mt, Carney:2017jut}. 
The limit $\lambda\to 0$ makes the amplitude vanish.

Because the first factor in \eqref{D} is independent of photon states, we always have the vanishing factor
$\left(\frac{\lambda}{\Lambda}\right)^A$ in any transition amplitudes  unless states have something cancelling it. 
Thus, transition amplitudes 
$\bra{\gamma_1}e^{\Delta R}\ket{\gamma_2}$ between arbitrary Fock states of hard photons $\ket{\gamma_{1,2}}$ vanish in the limit $\lambda\to 0$. 
This is not the case for dressed states including soft photons as we will see in section~\ref{sec:IRfinite}.

The result that there is no transition between Fock states is actually the conservation law of the asymptotic symmetry in QED as pointed out in \cite{Kapec:2017tkm}. 
In a scattering process, the asymptotic symmetry requires that  the initial and final states belong to different soft-charge sectors.\footnote{More precisely, the initial and final states must have the same asymptotic charge $Q_{as}=Q_{soft}+Q_{hard}$. The hard charge $Q_{hard}$ generally changes in the scattering, and thus the soft charge $Q_{soft}$ also has to change. }
Since the Fock states belong to the same sector, the transition between them are not allowed.  
It means that we have to dress the initial or final states so that it is consistent with the asymptotic symmetry. 
We will show this in section~\ref{sec:AS} for the background current model \eqref{eq:Lag}. 

In the above discussions, we have ignored $R_{out}(t_f)$ and  $R_{in}(t_i)$ in $e^{R_{out}(t_f)+\Delta R-R_{in}(t_i)}$ for simplicity. 
One might think that these extra  operators cancel the IR divergences in \eqref{ampIRdiv}.  
Such cancellation does not occur. 
In \cite{Kulish:1970ut} and the literature, it is discussed that the oscillating factors $e^{\pm i\frac{p_{n} \cdot k}{E_{p_n}}t}$ in \eqref{def:Rf} and \eqref{def:Ri} can be set to 1 because we consider the IR region $|\vk|\ll 1$. 
If this is correct, we have $R_{out}(t_f)\sim R_{out}(0)$ and $R_{out}(t_i)\sim R_{in}(0)$, and there is no IR divergence in $e^{R_{out}(t_f))+\Delta R-R_{in}(t_i)}$ because
\begin{align}
e^{R_{out}(t_f)+\Delta R-R_{in}(t_i)}=e^{R_{out}(t_f)-R_{out}(0)+R_{in}(0)-R_{in}(t_i)}\sim e^{0}\,.
\end{align}
However, we cannot use this approximation because it is only valid for the parameter region,
$
\left|\frac{p\cdot k}{E_{p}}\right|\ll \left|\frac{1}{t}\right|\,,
$
and this region vanishes when we take the limit $t\rightarrow \pm \infty$. 
In general, $e^{R_{out}(t_f)-R_{in}(t_i)}$ produces IR divergences oscillating  with $t_f, t_i$ which are not the same as \eqref{ampIRdiv}. 
Nevertheless, we do not need to worry much about these divergences from $R_{out}(t_f)$ and $R_{in}(t_i)$. 
As we will see in next section, the gauge invariance requires the asymptotic states have the dressing factors $e^{-R_{out}(t_f)}$, $e^{R_{in}(t_i)}$ which almost cancel $R_{out}(t_f)$ and $R_{in}(t_i)$ coming from the time-evolution operator \eqref{teo}. 
After this cancellation, we still have the IR divergence given by \eqref{ampIRdiv}.\footnote{Some problems on the Kulish-Faddeev dressed sates are also discussed for a massless Yukawa model in  \cite{Dybalski:2017mip}.}  
This is the reason  why we need other dressings in addition to $e^{-R_{out}(t_f)}$, $e^{R_{in}(t_i)}$.

\subsection{Gauge invariant states are dressed states \label{sec:Gauss}}
The initial and final states should be physical states. 
In the BRST formalism, the physical state condition is given by the BRST closed condition \cite{Kugo:1977zq},
\begin{align}
\label{BRST-cond}
    Q_{BRST}\ket{phys}=0.
\end{align}
It was shown in \cite{Hirai:2019gio} that the physical states cannot be the Fock states if there are interactions. 
This fact just represents that Coulomb-like fields must exist around charged particles. 
This fact also holds in the background current model \eqref{eq:Lag}. 
The discussion is slightly different from the full QED because of the explicit time-dependence of the background current. 
Thus, we briefly review the analysis of the physical state condition \eqref{BRST-cond} in \cite{Hirai:2019gio} for the background current model \eqref{eq:Lag}. 

The BRST charge in the Schr\"{o}dinger picture is given by 
\begin{align}
\label{eq:BRSTop}
Q_{BRST}^\mathrm{s}(t)=-\intkw \left[
c(\vk) \{k^\mu a_\mu^{\dagger}(\vk) +e^{-i\omega t_s}j^{0}_{\mathrm{pp}}(t,-\vk)\}
+c^\dagger(\vk)  \{k^\mu a_\mu(\vk) +e^{i\omega t_s}j^{0}_{\mathrm{pp}}(t,\vk)\}
\right],
\end{align}
where $c(\vk), c^\dagger(\vk)$ are annihilation and creation operators of the ghost field. We summarize the BRST formalism in appendix~\ref{app:BRST}.

The physical Hilbert space is given by the BRST cohomology of
this BRST charge. 
However,  
there is a tricky point because 
the BRST charge \eqref{eq:BRSTop} does not commute with the Hamiltonian unlike the full QED case (see appendix~\ref{app:BRST}). 
Thus, one might worry that the BRST closed condition is not consistent with the time-evolution. 
This is not the case. 
The BRST charge \eqref{eq:BRSTop} has an explicit time-dependence through the background current $j^{\mu}_{\mathrm{pp}}$,  and it helps the consistency.  
One can show (see appendix~\ref{app:BRST} for the proof) that if a state $\ket{\psi(t)}_s$ is a BRST closed state at time $t$ as  $Q^\mathrm{s}_{BRST}(t)\ket{\psi(t)}_s =0$, the time-evolved state  
$U(t',t)\ket{\psi(t)}_s$ is also a BRST closed state at time $t'$ as $Q^\mathrm{s}_{BRST}(t')U(t',t)\ket{\psi(t)}_s =0$.

We now solve the BRST closed condition $Q^\mathrm{s}_{BRST}(t)\ket{\psi(t)}_s =0$. 
We restrict states to the ghost vacuum as $c(\vk)\ket{\psi(t)}_s=0$ since the ghost is decoupled from the dynamics of gauge fields. 
The condition $Q^\mathrm{s}_{BRST}(t)\ket{\psi(t)}_s =0$ then becomes 
\begin{align}
\label{Gauss-law}
     \left[k^\mu a_\mu(\vk) +e^{i\omega t_s}j^{0}_{\mathrm{pp}}(t,\vk)\right] \ket{\psi(t)}_s=0
\end{align}
for any $\vk$. 
The key point is that this condition is different from the conventional Gupta-Bleuler condition $k^\mu a_\mu(\vk) \ket{\psi(t)}_s=0$. 
The form of \eqref{Gauss-law} means that physical states are coherent states. 
Non-zero charges must be surrounded by an infinite number of (longitudinal) photons, namely  Li\'enard-Wiechert potential.
In this sense, the states are not elements of the Fock space where the number of photons is finite. 

In the interaction picture, the condition \eqref{Gauss-law} is written as 
\begin{align}
\label{Gauss-law_int}
     \left[k^\mu a_\mu(\vk) +e^{i\omega t}j^{0}_{\mathrm{pp}}(t,\vk)\right] \ket{\psi(t)}_I=0\,.
\end{align}
The explicit form of the background current \eqref{j_cl} means 
\begin{align}
    j^{0}_{\mathrm{pp}}(t,\vk)=\Theta(-t)\sum_{n \in I} e_n\,e^{-i \frac{\vp_n\cdot \vk }{E_n}t} 
+\Theta(t)\sum_{n \in  F} e_{n}\,e^{-i \frac{\vp_{n}\cdot \vk }{E_n}t}.
\end{align}
Using this, we can solve \eqref{Gauss-law_int} as\footnote{We suppose $\Theta(0)=1/2$. It implies  that $\ket{\psi(t=0)}_I=e^{\frac{1}{2}\left(R_{in}(0)+R_{out}(0)\right)}\ket{\psi}_0$.}
\begin{align}
\label{FK-dress-st}
   \ket{\psi(t)}_I=  \left\{ \begin{array}{ll}
e^{R_{in}(t)}\ket{\psi}_0 & (t<0) \\
e^{R_{out}(t)}\ket{\psi}_0 & (t>0)
\end{array} \right.,
\end{align}
where $\ket{\psi}_0$ are arbitrary states satisfying the conventional  Gupta-Bleuler condition $k^\mu a_\mu(\vk) \ket{\psi}_0=0$, and $R_{out}$ and $R_{in}$ are given in \eqref{def:Rf}, \eqref{def:Ri}.  
These $R_{out}$ and $R_{in}$ are nothing but the Kulish-Faddeev 
dressing operators \cite{Kulish:1970ut}. 
As mentioned above, 
the condition \eqref{Gauss-law} is Gauss's law.
The dressing factors in \eqref{FK-dress-st} actually correspond to the Li\'enard-Wiechert potential
\cite{Bagan:1999jf, Hirai:2019gio}.

An important remark is that the states $\ket{\psi}_0$ in \eqref{FK-dress-st} are not necessarily the Fock states.  
$\ket{\psi}_0$ also can be dressed states if they satisfy the condition $k^\mu a_\mu(\vk) \ket{\psi}_0=0$. In this sense, the dressing factors are not uniquely fixed by Gauss's law condition \eqref{Gauss-law_int}. 
It was argued in \cite{Hirai:2019gio} that this freedom of the dressing factors is important in the conservation law of asymptotic symmetry.
We will see that we need dressing factors other than $R_{out}(t), R_{in}(t)$ to have IR finite $S$-matrix elements.

\subsection{Asymptotic symmetry requires a new dress}\label{sec:AS}
Gauge theories on manifolds with asymptotic boundaries generally have asymptotic symmetries. 
For QED on 4d Minkowski space, the asymptotic symmetry consists of large U(1) gauge transformations \cite{He:2014cra, Campiglia:2015qka, Kapec:2015ena}.
Although the asymptotic symmetry is a part of local U(1) symmetry, it is not a gauge redundancy.\footnote{See \cite{Hirai:2018ijc} for an explanation in the BRST formalism.}
In general, a physical symmetry leads to a conservation law. 
For example, total electric charges must conserve in any scattering process. 
This is the global U(1) charge conservation. 
Similarly, we have an infinite number of conservation laws associated with a part of large U(1) gauge transformations. 
The relation between the asymptotic symmetry and dressed states was discussed, \textit{e.g.}, in \cite{Mirbabayi:2016axw, Gabai:2016kuf, Kapec:2017tkm,  Carney:2018ygh, Hirai:2019gio}. 
In \cite{Hirai:2019gio}, it was argued that the Kulish-Faddeev dressing operator is not enough to realize the conservation laws of the asymptotic symmetry.  
In this section we will see that the asymptotic symmetry requires additional dressing factors in order to have non-vanishing $S$-matrix elements.

The background current model \eqref{eq:Lag} with the Feynman gauge is invariant under the residual gauge symmetry $A_\mu \to A_\mu+\partial_\mu \epsilon$ where the gauge parameter $\epsilon$ satisfies $\Box \epsilon=0$. 
The asymptotic symmetry is then given by these transformations such that the parameters $\epsilon(x)$ are $\mathcal{O}(1)$ at the asymptotic boundary. 
We can reach the future null infinity $\mathscr{I}^+$ by taking $t, r \to +\infty$ with the retarded time $u=t-r$ fixed.
$\mathscr{I}^+$ is parameterized by $u$ and angular coordinates $\Omega^A\, (A=1,2)$. 
By solving $\Box \epsilon=0$ near the future null infinity $\mathscr{I}^+$,  we can find that the asymptotic behavior of $\epsilon(x)$ at $\mathscr{I}^+$ is given by an arbitrary function of the angular coordinates $\epsilon^{(0)}(\Omega)$.
If we specify this boundary behavior $\epsilon^{(0)}(\Omega)$, the bulk parameter $\epsilon(x)$ is fixed up to an addition of ``small" gauge parameter which is a gauge redundancy. 
Thus, the asymptotic symmetry is an infinite dimensional symmetry parameterized by  functions $\epsilon^{(0)}(\Omega)$ on the unit two-sphere. 
The expression of $\epsilon(x)$ as a functional of $\epsilon^{(0)}(\Omega)$ is given by \eqref{LGP} in appendix~\ref{app:as}.

In the Schr\"{o}dinger picture, the Noether charge for the gauge transformation is given by 
\begin{align}
Q_{as}^\mathrm{s}[\epsilon]=\intx \left[
-\Pi^{0s} \partial_0 \epsilon -\Pi^{is}\partial_i \epsilon
+j^0_{\mathrm{pp}}\epsilon
\right].
\label{neother_as}
\end{align}
Note that, in order to match the convention with that in the full QED, we have added the last term $j^0_{\mathrm{pp}}\epsilon$, which is a trivial operator (c-number) because $j^\mu_{\mathrm{pp}}$ is a classical background current.
If $\epsilon$ vanishes at the asymptotic boundary, the charge \eqref{neother_as} is BRST exact\footnote{This charge can be written as $Q_{as}^\mathrm{s}[\epsilon]=-\intx\, \partial_i (\Pi^{is}\epsilon)
+ \left\{Q^\mathrm{s}_{BRST}, \intx(-\bar{c}^\mathrm{s} \partial_0 \epsilon+i \pc^\mathrm{s} \epsilon)
\right\}$ where $\bar{c}^\mathrm{s}, \pc^\mathrm{s}$ are ghost operators (see appendix~\ref{app:BRST}).} and does not play any role on the physical Hilbert space. 
It just means that $Q_{as}^\mathrm{s}[\epsilon]$ for  a ``small" gauge parameter $\epsilon$ generates a gauge redundancy which is not  physical symmetry. 
If $\epsilon$ is a  ``large" gauge parameter as discussed above,  $Q_{as}^\mathrm{s}[\epsilon]$ is a charge for the asymptotic symmetry, which we call the asymptotic charge.  
It can act nontrivially on the physical Hilbert space unlike small gauge charges.
We can classify physical states by the values of the asymptotic charges because
  $Q_{as}^\mathrm{s}[\epsilon]$ are commutative $[Q_{as}^\mathrm{s}[\epsilon], Q_{as}^\mathrm{s}[\epsilon']]=0$ and also commute with the BRST charge $[Q_{as}^\mathrm{s}[\epsilon],Q^\mathrm{s}_{BRST}(t)]=0$.

We now consider a time-evolution of an initial state $\ket{\beta(t_i)}_s$, which is converted to the interaction picture as $\ket{\beta(t_i)}_s=U_0(t_i,t_s)\ket{\beta(t_i)}_I$ where $\ket{\beta(t_i)}_I$ should be dressed as $e^{R_{in}(t_i)}\ket{\beta}_0$ (see $\eqref{FK-dress-st}$).
For simplicity, we assume that $\ket{\beta(t_i)}_I$ is an eigenstate of $Q_{as}^I[\epsilon(t_i)]$.\footnote{General states can be obtained by the superposition of the eigenstates.}
It means that $\ket{\beta}_0$ is also an eigenstate of $Q_{as}^I[\epsilon(t_i)]$ because the commutator of $Q_{as}^I[\epsilon(t_i)]$ and the dressing factor $R_{in}(t_i)$ is a c-number. 
Supposing that the eigenvalue for $\ket{\beta}_0$ is $Q_{\beta}[\epsilon(t_i)]$, we have
\begin{align}
\label{Q^Ii-ei}
  Q_{as}^I[\epsilon(t_i)]\ket{\beta(t_i)}_I
  =\left([Q_{as}^I[\epsilon(t_i)],R_{in}(t_i)]+Q_\beta[\epsilon(t_i)]\right)\ket{\beta(t_i)}_I .
\end{align}
We can also find that a time-evolved state $\ket{\beta(t_f)}_I$ is an eigenstate of $Q_{as}^I[\epsilon(t_i)]$ (not $Q_{as}^I[\epsilon(t_f)]$) as follows.
From \eqref{teo}, $\ket{\beta(t_f)}_I$ can be written as
\begin{align}
\ket{\beta(t_f)}_I
=U^{-1}_{0}(t_f,t_s)U(t_f,t_i)\ket{\beta(t_i)}_s
=e^{i\Phi(t_f,t_i)}e^{\tilde{R}}  \ket{\beta(t_i)}_I,
\end{align}
where 
\begin{align}\label{eq:Rtilde}
\tilde{R}:=R_{out}(t_f)+\Delta R-R_{in}(t_i).
\end{align}
Using the expression, we obtain 
\begin{align}
 Q_{as}^I[\epsilon(t_i)]\ket{\beta(t_f)}_I
 &=e^{i\Phi(t_f,t_i)}Q_{as}^I[\epsilon(t_i)]e^{\tilde{R}}  \ket{\beta(t_i)}_I \non
 &=e^{i\Phi(t_f,t_i)}\left([Q_{as}^I[\epsilon(t_i)], e^{\tilde{R}}]+  e^{\tilde{R}}Q_{as}^I[\epsilon(t_i)]\right)\ket{\beta(t_i)}_I\non
 &=\left([Q_{as}^I[\epsilon(t_i)], \tilde{R}]+[Q_{as}^I[\epsilon(t_i)], R_{in}(t_i)]+Q_\beta[\epsilon(t_i)]\right)\ket{\beta(t_f)}_I,
 \label{eq:Qieigen}
\end{align}
where we have used the fact that 
$[Q_{as}^I[\epsilon(t_i)], \tilde{R}]$ is a c-number and eq.\eqref{Q^Ii-ei} in the third equality.
Putting \eqref{eq:Rtilde} into \eqref{eq:Qieigen}, we obtain
\begin{align}
Q_{as}^I[\epsilon(t_i)]\ket{\beta(t_f)}_I=\left(\,[Q_{as}^I[\epsilon(t_i)], R_{out}(t_f)]+\Delta Q+Q_\beta[\epsilon(t_i)]\,\right)\ket{\beta(t_f)}_I ,
\end{align}
where 
\begin{align}
\Delta Q:=[Q_{as}^I[\epsilon(t_i)], \Delta R].
\end{align}
Therefore, $\ket{\beta(t_f)}_I$ is the eigenstate of $Q_{as}^I[\epsilon(t_i)]$ with the eigenvalue $[Q_{as}^I[\epsilon(t_i)], R_{out}(t_f)]+\Delta Q+Q_\beta[\epsilon(t_i)]$.
This is an important result for finding the appropriate asymptotic state for the following reason.
    
The amplitude in \eqref{eq:defSmatrix}
\begin{align}
S_{\alpha,\beta}=\tensor[_s]{\bra{\alpha(t_f)}}{}U(t_f,t_i)\ket{\beta(t_i)}_s
=\tensor[_I]{\bra{\alpha(t_f)}\ket{\beta(t_f)}}{_I}
\end{align}
 vanishes unless $\ket{\alpha(t_f)}_I$ has the same eigenvalue $[Q_{as}^I[\epsilon(t_i)], R_{out}(t_f)]+\Delta Q+Q_\beta[\epsilon(t_i)]$ of $Q_{as}^I[\epsilon(t_i)]$ as $\ket{\beta(t_f)}_I$. 
On the other hand, the gauge invariance requires that  $\ket{\alpha(t_f)}_I$ be dressed as $\ket{\alpha(t_f)}_I=e^{R_{out}(t_f)}\ket{\alpha}_0$.
It means that, in order to have a non-zero amplitude $S_{\alpha,\beta}$, the state $\ket{\alpha}_0$ must satisfy the following eigenstate equation:   
\begin{align}
    Q_{as}^I[\epsilon(t_i)] \ket{\alpha}_0
    =\left(\Delta Q+Q_\beta[\epsilon(t_i)]\right)\ket{\alpha}_0\,.
    \label{change_Qas}
\end{align}
Here let us represent the Hilbert space satisfying the conventional Gupta-Bleuler condition by $\mathcal{H}^0$, 
\begin{align}
\ket{\phi}\in \mathcal{H}^0\ \Leftrightarrow \ k^\mu a_{\mu}(k)\ket{\phi}=0\quad \text{for any $k^\mu$},
\end{align}
and the subspace by $\mathcal{H}^0_{Q}$  where the eigenvalue of $Q_{as}^I[\epsilon(t_i)]$ is $Q[\epsilon(t_i)]$. 
Then the above  says that  $\ket{\alpha}_0$ and  $\ket{\beta}_0$ need to belong to different sectors with respect to the asymptotic charges $Q_{as}^I[\epsilon(t_i)]$ as
\ba{
\label{eq:shift_ab}
\ket{\beta}_0 \in\mathcal{H}^0_{Q}\ \Rightarrow \ \ket{\alpha}_0\in\mathcal{H}^0_{Q+\Delta Q}\,,
}
in order to have the non-zero matrix element $S_{\alpha,\beta}$.
We need a dressing factor associated to the shift $\Delta Q$ in $\ket{\alpha}_0$. 
This transition can be realized by adding the transverse part of the dressing operator $\Delta R$ as follows.

We first introduce our notation of the polarization vectors. 
We represent two transverse polarization vectors by $\epsilon_A^\mu(\hat{k})=(0,\vec{\epsilon}_A(\hat{k}))$ ($A=1,2$) where $\vec{\epsilon}_A$ depends only on the direction of photon momentum $\vk$ and  satisfies $\vk \cdot \vec{\epsilon}_A(\hat{k})=0$. 
Here, we take a real basis such that $\epsilon^{\mu}_A=\epsilon^{\mu\ast}_A$. 
We also represent the inner product of two polarization vectors by $\eta^{T}_{AB}$
as $\eta^{T}_{AB}(\hat{k})=\eta_{\mu\nu}\epsilon_A^{\mu}(\hat{k})\epsilon_B^\nu(\hat{k})$.
This matrix $\eta^{T}_{AB}$ must be  invertible so that the two polarization vectors are independent. 
We have the completeness relation 
\begin{align}
    \eta^{\mu\nu}=\eta^{AB}_{T}\epsilon_A^\mu(\hat{k})\epsilon^{\nu}_B(\hat{k})
    -\frac{k^\mu \tilde{k}^\nu+\tilde{k}^\mu k^\nu}{2\omega^2},
\end{align}
where $\eta^{AB}_{T}$ is the inverse matrix of $\eta^{T}_{AB}$ as $\eta^{AB}_{T}\eta^{T}_{BC}=\delta^A_C$, and  $\tilde{k}^\mu$ is the spatially reflected vector of $k^\mu$ as  $\tilde{k}^\mu=(\omega, -\vk)$ for $k^\mu=(\omega, \vk)$.
Then, we can decompose the dressing operator $\Delta R$ into the transverse and longitudinal parts as 
\begin{align}
\Delta R=  -R_{out}(0)+R_{in}(0)=\Delta R_{T}  +\Delta R_{L} 
\end{align}
with 
\begin{align}
\label{def:RT}
    \Delta R_{T}&=-\sum_{n}\intkw \eta^{AB}_{T}(\hat{k})\frac{\eta_n e_{n} p_{n}\cdot \epsilon_A(\hat{k})}{p_{n}\cdot k} \left[\epsilon_B(\hat{k}) \cdot a(\vk)
    -\epsilon_B(\hat{k}) \cdot a^{\dagger}(\vk)
    \right],
    \\
    \label{def:RL}
    \Delta R_{L}&=\sum_{n}\int \frac{\widetilde{d^3 k}}{2\omega^2}
    \eta_n e_{n} \left[\tilde{k} \cdot a(\vk)
    -\tilde{k} \cdot a^{\dagger}(\vk)
    \right]
    +
    \sum_{n}\int  \frac{\widetilde{d^3 k}}{2\omega^2} \frac{\eta_n e_{n} p_{n}\cdot \tilde{k}}{p_{n}\cdot k} \left[k \cdot a(\vk)
    -k \cdot a^{\dagger}(\vk)
    \right]
    \nn
    &= \sum_{n}\int \frac{\widetilde{d^3 k}}{2\omega^2} \frac{\eta_n e_{n} p_{n}\cdot \tilde{k}}{p_{n}\cdot k} \left[k \cdot a(\vk)
    -k \cdot a^{\dagger}(\vk)
    \right],
\end{align}
where we have used the total electric charge conservation $\sum_n \eta_n e_n=0$.
One can easily show that the longitudinal part $\Delta R_{L}$ commutes with $\Pi^I_\mu$. 
It means that $\Delta R_{L}$ commutes with the asymptotic charges $Q_{as}^I[\epsilon]=\intx \left(
-\Pi^{0I} \partial_0 \epsilon -\Pi^{iI}\partial_i \epsilon
+j^0_\mathrm{pp}\epsilon
\right)$. 
Thus, the shift of the asymptotic charges $\Delta Q=[ Q_{as}^I[\epsilon(t_i)], \Delta R]$, which appeared in \eqref{change_Qas}, can be written as
\begin{align}
    \Delta Q=[ Q_{as}^I[\epsilon(t_i)], \Delta R_{T}].
\end{align}
It shows that the dressing factor $e^{\Delta R_{T}}$ shifts the asymptotic charges by $\Delta Q$. 
If a state $\ket{\psi}_0$ satisfies the conventional Gupta-Bleuler condition, the dressed state $e^{\Delta R_{T}}\ket{\psi}_0$ also does, because $\Delta R_{T}$ consists only of transverse components.   
Therefore, the transition from $\mathcal{H}^0_{Q}$ to $\mathcal{H}^0_{Q+\Delta Q}$ can be implemented by the dressing factor $e^{\Delta R_{T}}$ as
\begin{align} 
\mathcal{H}^0_{Q+\Delta Q}=e^{\Delta R_{T}}\mathcal{H}^0_{Q}. 
\end{align}
Note that each term in the sum in $\Delta R_{T}$ is the same as  Chung's dressing operator which is acted on each charged particle to cancel IR divergences in \cite{Chung:1965zza}.
We will see in section \ref{sec:IRfinite} that this additional factor $e^{\Delta R_{T}}$ makes the $S$-matrix IR finite.

\subsubsection{Asymptotic charge conservation\label{sec:asyCL}}
In this subsection, we comment on the conservation law of the asymptotic symmetry.\footnote{You may skip this part if you would like to see soon the IR finiteness of the $S$-matrix.}

We have shown that $\ket{\alpha}_0$ and  $\ket{\beta}_0$ belong to different sectors with respect to the asymptotic charges $Q_{as}^I[\epsilon(t_i)]$ as \eqref{eq:shift_ab}.
The point is that we considered  $Q_{as}^I[\epsilon(t_i)]$, not $Q_{as}^I[\epsilon(t_f)]$, even for the final state $\ket{\alpha}_0$. 
It is difficult to find the relation between $Q_{as}^I[\epsilon(t_i)]$ and $Q_{as}^I[\epsilon(t_f)]$ for finite $t_i$ and  $t_f$.
However, a simplification occurs in the asymptotic limit
$t_f\to \infty, t_i\to -\infty$.
We will show that, if the initial state $\ket{\beta}_0$ is an eigenstate of the past charge $\lim_{t_i\to -\infty}Q_{as}^I[\epsilon(t_i)]$ with the eigenvalue $Q[\epsilon^{(0)}]$, the final state $\ket{\alpha}_0$ is an eigenstate of the future charge $\lim_{t_f\to \infty}Q_{as}^I[\epsilon(t_f)]$ with the same eigenvalue $Q[\epsilon^{(0)}]$ in order to have the nonzero amplitude $\tensor[_s]{\bra{\alpha(t_f)}}{}U(t_f,t_i)\ket{\beta(t_i)}_s$. 

We first consider the asymptotic limits of the asymptotic charge, $\lim_{t\to \pm\infty} Q_{as}^I[\epsilon(t)]$. 
The gauge parameter $\epsilon(x)$ approaches a function on the celestial sphere, $\epsilon^{(0)}(\Omega)$, near the future null infinity $\mathscr{I}^+$.
As will be shown in appendix~\ref{app:as}, the limits of the asymptotic charge can be computed as
\begin{align}
\label{eq:as_consv}
&Q_{as}^{-I}[\epsilon^{(0)}]
:=\lim_{t_i\to -\infty} Q_{as}^I[\epsilon(t_i)]=Q_{hard}^{in}[\epsilon^{(0)}]+Q_{soft}[\epsilon^{(0)}]\,,\non
&Q_{as}^{+I}[\epsilon^{(0)}]
:=\lim_{t_f\to +\infty} Q_{as}^I[\epsilon(t_f)]=Q_{hard}^{out}[\epsilon^{(0)}]+Q_{soft}[\epsilon^{(0)}],
\end{align}
where 
\begin{align}
 Q_{hard}^{out\,/\,in}[\epsilon^{(0)}]
 &:=\sum_{n \in F\, /\,I} 
     \int d^2 \Omega  \sqrt{\gamma}
   \frac{e_{n} m_{n}^2  \epsilon^{(0)}(\Omega)}{4\pi(-E_{n}+\vp_{n} \cdot \hat{x}(\Omega))^2},\\
Q_{soft}[\epsilon^{(0)}]&:=\frac{1}{8\pi}\lim_{\omega \to 0} \omega
  \int d^2\Omega \sqrt{\gamma}\gamma^{AB}\partial_A  \hat{x}^i\partial_B \epsilon^{(0)}
\left[ a_i(\omega\hat{x}) +  a_i^{\dagger}(\omega\hat{x})\right].
\end{align}
The hard charges $Q_{hard}^{out/in}$ are c-numbers which agree with those computed in the classical case in \cite{Hirai:2018ijc}.
$Q_{soft}$ is an operator acting on the soft photon sector.

In the asymptotic limit, $\Delta Q$ appeared in \eqref{change_Qas} is also simplified as 
\begin{align}
    \lim_{t_i \to -\infty}\Delta Q&=[Q_{soft}, \Delta R]
   =-Q_{hard}^{out}+Q_{hard}^{in}.
\end{align}

We suppose that the initial state $\ket{\beta}_0$ is an eigenstate of $Q_{as}^{-I}[\epsilon^{(0)}]$ with the eigenvalue $Q[\epsilon^{(0)}]$.
Then, as shown by eq.\eqref{change_Qas}, 
in order to have the nonzero amplitude $\tensor[_s]{\bra{\alpha(\infty)}}{}U(\infty,-\infty)\ket{\beta(-\infty)}_s$, 
the final state $\ket{\alpha}_0$ must be the eigenstate of the past charge $Q_{as}^{-I}[\epsilon^{(0)}]$ with the eigenvalue 
\begin{align}
    Q[\epsilon^{(0)}]+\lim_{t_i\to -\infty}\Delta Q
    = Q[\epsilon^{(0)}]-Q_{hard}^{out}+Q_{hard}^{in}.
\end{align}
Combining this equation and eq.\eqref{eq:as_consv},  we find that 
$\ket{\alpha}_0$ is the eigenstate of the future charge 
$Q_{as}^{+I}[\epsilon^{(0)}]$ with the eigenvalue $Q[\epsilon^{(0)}]$. 

The above properties of the eigenvalues of the asymptotic charges mean the asymptotic charge conservation, that is, the equation 
\begin{align}
\label{as_cnsv_I}
\tensor[_I]{\bra{\alpha(\infty)}}{}
Q_{as}^{+I}[\epsilon^{(0)}]
S_0(\infty,-\infty)\ket{\beta(-\infty)}_I
=\tensor[_I]{\bra{\alpha(\infty)}}{}
S_0(\infty,-\infty)
Q_{as}^{-I}[\epsilon^{(0)}]
\ket{\beta(-\infty)}_I
\end{align}
holds if the amplitude $\tensor[_I]{\bra{\alpha(\infty)}}{}S_{0}(\infty,-\infty)\ket{\beta(-\infty)}_I$ is nonzero.
Since $\ket{\beta(t_i)}_I=e^{R_{in}(t_i)}\ket{\beta}_0$,
the initial state
$\ket{\beta(t_i)}_I$ is the eigenstate of $Q_{as}^I[\epsilon(t_i)]$ with the eigenvalue
\begin{align}
    Q[\epsilon(t_i)]+[Q_{as}^I[\epsilon(t_i)],R_{in}(t_i)]
\end{align}
if $\ket{\beta}_0$ has the eigenvalue $Q[\epsilon(t_i)]$.
The commutator $[Q_{as}^I[\epsilon(t_i)],R_{in}(t_i)]$ is computed as
\begin{align}\label{eq:chargeofLW}
    [Q_{as}^I[\epsilon(t_i)],R_{in}(t_i)]=\sum_{n\in I}\int\!\! d^3 x F_{0i}^{LW}(t_i,\vx;p_n)\partial^i \epsilon(t_i,\vx), 
\end{align}
where $F_{0i}^{LW}(x;p_n)$ represents the classical Coulomb field for the Li\'enard-Wiechert potential of a uniformly moving charged particle with momentum $p_n$, 
\begin{align}
  F_{0i}^{LW}(x;p_n)
  :=i e_n\intkw \frac{ E_{n}k_i-p_{ni}\omega }{p_{n}\cdot k} \left(e^{-i\vk\cdot\left(\vx-\frac{\vp_{n} t_i}{E_{n}}\right)}-e^{i\vk\cdot\left(\vx-\frac{\vp_{n} t_i}{E_{n}}\right)}\right).  
\end{align}
The charge \eqref{eq:chargeofLW}  represent the contribution of the Coulomb field to  the asymptotic charge. It vanishes at asymptotic infinities because the Coulomb field  falls off at asymptotic infinity (see appendix \ref{app:falloff} for more 
detailed analysis.).
Thus, $\ket{\beta(-\infty)}_I$ has the same eigenvalue of $Q_{as}^{-I}[\epsilon^{(0)}]$ as $\ket{\beta}_0$. 
Similarly, 
$\ket{\alpha(\infty)}_I$ has the same eigenvalue of $Q_{as}^{+I}[\epsilon^{(0)}]$ as $\ket{\alpha}_0$. 
Since the eigenvalue of $Q_{as}^{+I}[\epsilon^{(0)}]$ for $\ket{\alpha}_0$ agrees with that of $Q_{as}^{-I}[\epsilon^{(0)}]$ for $\ket{\beta}_0$ as shown above, we obtain the asymptotic conservation law \eqref{as_cnsv_I}.
This conservation law can be symbolically written as 
\begin{align}
    Q_{soft}+Q_{hard}^{out}=Q_{soft}+Q_{hard}^{in}.
\end{align}
This is the quantum analog of the electromagnetic memory effect \cite{Bieri:2013hqa, Susskind:2015hpa, Hirai:2018ijc} (see also section~4 in \cite{Hirai:2019gio}).

\subsection{IR finiteness of dressed $S$-matrix}
\label{sec:IRfinite}
We now reconsider the transition amplitude $S_{\alpha,\beta}$ in \eqref{eq:defSmatrix} and see that there is no IR divergences for appropriate dressed states.
As shown in section~\ref{sec:Gauss}, the gauge invariance requires that 
the initial and final states be dressed states. 
In the interaction picture, the initial states $\ket{\beta(t_i)}_I$ are dressed as $\ket{\beta(t_i)}_I=e^{R_{in}(t_i)}\ket{\beta}_0$.
The state  $\ket{\beta}_0$ is an element of $\mathcal{H}^0$ subject to the conventional Gupta-Bleuler condition $k^\mu a_\mu(\vk)\ket{\beta}_0=0$.
The space $\mathcal{H}^0$ can be classified by the eigenvalues of the asymptotic charges $Q_{as}^I[\epsilon(t_i)]$ as $\mathcal{H}^0=\oplus_{Q} \mathcal{H}^0_Q$. 
We focus on the case where the state $\ket{\beta}_0$ has a definite eigenvalue $Q$, \textit{i.e.} $\ket{\beta}_0 \in \mathcal{H}^0_Q$, since arbitrary states in $\mathcal{H}^0$ can be written as superposition of eigenstates of $Q_{as}^I[\epsilon(t_i)]$.
As we showed in section~\ref{sec:AS}, 
the final state has to have an additional dressing factor to realize the shift of the asymptotic charges. 
We dress the final state as
\begin{align}
    \ket{\alpha(t_f); \Delta R_T}_I=e^{R_{out}(t_f)}e^{\Delta R_T}\ket{\alpha}_0 \qquad \text{with} \quad \ket{\alpha}_0 \in \mathcal{H}^0_Q,
\end{align}
where we write $\Delta R_T$ in the ket to stress that the dressing is different from $\ket{\beta(t_i)}_I$ by the additional dressing factor $e^{\Delta R_T}$.\footnote{Note that $\ket{\alpha}_0$ is different from that in section~\ref{sec:AS} where $\ket{\alpha}_0 \in \mathcal{H}^0_{Q+\Delta Q}$.} 
Then, 
the amplitude $S_{\alpha,\beta}$ turns out to be diagonal as
\begin{align}
S_{\alpha,\beta}
&=\tensor[_s]{\bra{\alpha(t_f);\Delta R_{T} }}{}U(t_f,t_i)\ket{\beta(t_i) }_s\nn
&=e^{i\Phi(t_f,t_i)}\,\tensor[_0]{\bra{\alpha}}{}
e^{-\Delta R_T}e^{-R_{out}(t_f)}
e^{R_{out}(t_f)+\Delta R-R_{in}(t_i)}
e^{R_{in}(t_i)}\ket{\beta}_0
\nn
&=
e^{i\Phi(t_f,t_i)}e^{-\frac{1}{2}[R_{out}(t_f)-R_{in}(t_i),\, \Delta R ]+\frac{1}{2}[R_{out}(t_f), R_{in}(t_i)]}
\tensor[_0]{\bra{\alpha}}{}
e^{\Delta R_L}\ket{\beta}_0
\nn
&=
e^{i\tilde{\Phi}(t_f,t_i)}\delta_{\alpha,\beta},
\end{align}
where the phase $\tilde{\Phi}$ is a c-number defined by
\begin{align}
  i\tilde{\Phi}(t_f,t_i):= i\Phi(t_f,t_i)-\frac{1}{2}[R_{out}(t_f)-R_{in}(t_i), \Delta R]+\frac{1}{2}[R_{out}(t_f), R_{in}(t_i)],
\end{align}
which is independent of states $\alpha, \beta$.
Note that $\Delta R_L$ given by \eqref{def:RL} contains only longitudinal photons, and  thus we have $\tensor[_0]{\bra{\alpha}}{}e^{\Delta R_L}\ket{\beta}_0=\tensor[_0]{\braket{\alpha}{\beta}}{_0}=\delta_{\alpha,\beta}$.  

Combining the expression of $\Phi$ given by \eqref{eq:phase-int} with \eqref{eq:FKphase},
we can find that 
the total phase $\tilde{\Phi}$ is given by
\begin{align}
\label{eq:totalphase}
  \tilde{\Phi}(t_f,t_i)  = -\theta_{out}(t_f)+\theta_{in}(t_i)
\end{align}
where
\begin{align}
&\theta_{out}(t_f):=
\sum_{n\in F} \intkw \frac{e_{n}^2 m_n^2 }{(k \cdot p_n) E_n}t_f
+ \sum_{\substack{n,m \in F\\  n\neq m}}
\intkw\frac{e_{n}e_{m}  (v_{n}\cdot v_{m})}{(k \cdot v_{m})\, [k \cdot \pa{ v_{n}- v_{m}}]}\sin\pa{k \cdot \pa{ v_{n}- v_{m}}t_f},
\\
&\theta_{in}(t_i):=
\sum_{n\in I} \intkw \frac{e_{n}^2 m_n^2 }{(k \cdot p_n) E_n}t_I
+\sum_{\substack{n,m \in I\\  n\neq m}}
\intkw\frac{e_{n}e_{m}  (v_{n}\cdot v_{m})}{(k \cdot v_{m})\, [k \cdot \pa{ v_{n}- v_{m}}]}\sin\pa{k \cdot \pa{ v_{n}- v_{m}}t_i}.
\end{align}

The first terms in $\theta_{out}(t_f), \theta_{in}(t_i)$ are linear of $t_f, t_i$ respectively, and diverge in the limit $t_f\to \infty, t_i \to -\infty$. 
Because the phases are independent of photon states, we may also absorb the phases into the dressing factor as
\begin{align}
&\ket{\alpha(t_f);\Delta R_{T}}_I:=e^{R_{out}(t_f)}e^{\Delta R_T}e^{-i\theta_{out}(t_f)}\ket{\alpha}_0\,,\\ &\ket{\beta(t_i)}_I:=e^{R_{in}(t_i)}e^{-i\theta_{in}(t_i)}\ket{\beta}_0. 
\end{align}
Using these redefined dressed states, the $S$-matrix is trivial in the sense as
\begin{align}
S_{\alpha, \beta}
=\tensor[_s]{\bra{\alpha(t_f);\Delta R_{T} }}{}U(t_f,t_i)\ket{\beta(t_i) }_s
=\tensor[_0]{\braket{\alpha}{\beta}}{_0}
=\delta_{\alpha,\beta}\,.
\end{align}
Thus we have confirmed that the $S$-matrix for the dressed states is IR finite.
This result shows that the $S$-matrix in this model is trivial except for the dressing operators $R_{out}(t_f), R_{in}(t_i),  \Delta R_T$ and $e^{-i\theta_{out}(t_f)}, e^{-i\theta_{in}(t_i)}$. 
The first two operators $R_{out}(t_f), R_{in}(t_i)$ are necessary for Gauss's law constraint. 
$\Delta R_T$ is an operator realising the shift of the asymptotic charges, and $e^{-i\theta_{out}(t_f)}, e^{-i\theta_{in}(t_i)}$ are state-independent overall phases. 
This is consistent with the classical situation where the scattering of electromagnetic waves is trivial as discussed around eq.~\eqref{eq:cl_sol}.

\section{Discussion on full QED}\label{sec:fullQED}

We have seen that there are no IR divergences in the $S$-matrix elements in the background current model if we use the appropriate dressed states.
We will discuss that this is true even for the full QED.

As we saw, the naive IR divergences take  the same expressions \eqref{ampIRdiv} in the background current model and the full QED.
In fact, this result is roughly the soft photon theorem as follows. 
We define the classical current operator $j^\mu_{\mathrm{pp}}$ as
\begin{align}
\label{eq:j_cl_op}
  j^\mu_{\mathrm{pp}}(t,\vx)&:=  \sum_n e_n \intp \frac{p^\mu}{E_p}\, \delta^3(\vx -\vp t/E_p) \rho_n(\vp),
\\
\rho_n(\vp)&:=b_n^\dagger(\vp)b_n(\vp)-d_n^{\dagger}(\vp)d_n(\vp),
\end{align}
where $b_n^\dagger$ (and $d_n^{\dagger}$) are creation operators of the charged particles with charge $e_n$ (and anti-particles).\footnote{Here we omit labels for spins to simplify the expression. 
The normalization of the ladder operators is the same as \cite{Hirai:2018ijc, Hirai:2019gio}. 
$\rho_n$ represents  the number density of charge $e_n$.  }
The operator $j^\mu_{\mathrm{pp}}$ acts on the charged particle sectors as the classical current of point particles like \eqref{j_cl}.
In the Feynman diagrams for the full QED, 
the IR divergences arise when virtual soft photons interact with the external lines.
This interaction among soft photons and the on-shell charged particles is classical in the sense that we can replace the current operator $j^\mu$ with the above classical current operator $j^\mu_{\mathrm{pp}}$ as shown in the appendix~\ref{app:soft}. 
The proof is almost the same as that of soft photon theorem \cite{Weinberg:1965nx}.
Thus, when we evaluate IR divergences in the full QED, we can approximate the interaction $A_\mu j^\mu$ by $A
_\mu j^\mu_{\mathrm{pp}}$.
This is the reason why we have the same IR divergences in the background current model and the full QED. 

Since the structure of IR divergences are the same in the both model, 
the divergences in the full QED are canceled if we can use the same dressed states used in section~\ref{sec:IRfinite}.  
We need a clarification of the usage of these dressed states  because the physical state condition is now given by 
\begin{align}
\label{full-Gauss-law}
     \left[k^\mu a_\mu(\vk) +e^{i\omega t_s}j^{0}(t,\vk)\right] \ket{\psi(t)}_s=0,
\end{align}
which is different from \eqref{Gauss-law} in the current term.
Nevertheless, we can replace $j^{0}(t,\vk)$ by the classical current operator $j^{0}_{\mathrm{pp}}(t,\vk)$ in the asymptotic regions $t \sim \pm\infty$ using the saddle point approximation which becomes exact at $t = \pm\infty$ \cite{Hirai:2019gio}. 
Thus, we can use as the asymptotic states in the full QED the same dressed states used in the background current model. 
Then, there are no IR divergences in the $S$-matrix elements for these dressed states as we saw in the last section. 
Of course, the $S$-matrix in the full QED is non-trivial unlike the background current model because the interaction involving hard photons is not the same.
It would be an interesting future work to compute $S$-matrix elements for the full QED in our  dressed state formalism.

\section{Conclusion and outlook}\label{sec:conclusion}
We have shown that the Kulish-Faddeev dressed states are not enough to remove IR divergences in the $S$-matrix of QED. 
Although the Kulish-Faddeev dressed states, which were derived by solving the asymptotic dynamics in \cite{Kulish:1970ut}, are a solution of the physical state condition \eqref{Gauss-law_int} as eq.~\eqref{FK-dress-st}, 
it is able to add another dressing operator. 
The asymptotic symmetry actually requires us to add such another dressing operator at least to the initial or final state as we saw in section~\ref{sec:AS}. 
Reflecting this fact, if we use the original Kulish-Faddeev dressed states in both of the initial and final states, we encounter the IR divergences. 
In this paper, we suggest putting the additional dressing operator 
$e^{\Delta R_T}$ in addition to the original Kulish-Faddeev dressing operator $e^{R_{out}(t_f)}$ to the final state.\footnote{We also need to put some time-dependent but photon-state-independent phase factors to the initial and final states if we want to avoid an overall phase factor which infinitely oscillates in the asymptotic limit $t_i \to -\infty$, $t_f \to \infty$.}
For the new dressed states, IR divergences completely disappear, and the $S$-matrix is IR finite.  

The dressing operator $e^{\Delta R_T}$ is almost the same as that in Chung's paper \cite{Chung:1965zza}.
Chung's dressing is necessary to cancel IR divergences, but it is not compatible with the gauge invariance unless we add another dressing involving longitudinal photons. 
Kulish and Faddeev argued that their dressing is the same as Chung's dressing because the difference is just a factor like  $e^{i \frac{p\cdot k}{E}t}$ which becomes 1 for soft momenta $k \sim 0$ if $t$ is finite. 
However, we cannot use this approximation because  we take the limit $|t| \to \infty$.
Therefore, Kulish-Faddeev dressing is different from Chung's one. 
After all, we need both of them to cure IR divergences in a gauge invariant way, that is, our new dressed states are a mixture of the Chung and Kulish-Faddeev dressed states.  

Here we would like to stress that the conservation law of the asymptotic symmetry is a necessary condition to obtain the non-vanishing $S$-matrix elements, but it is not the sufficient condition.   
We introduced $\Delta R_T$ to realize the shift of the soft  charges like \eqref{change_Qas}. 
Just to realize this shift, we have many other possibilities. 
However, such dressings might not cancel the IR divergences.  

We also have a comment on the movability of the photon clouds discussed in \cite{Kapec:2017tkm, Choi:2017ylo}.
The dressing factor $e^{\Delta R_T}$ which we put to the final state in this paper can be moved to the initial state. 
Furthermore, we can decompose $\Delta R_T$ into two parts,  and put a part of $\Delta R_T$ to the final states and the rest to the initial state. 
Any decomposition is compatible with the gauge invariance and the asymptotic symmetry. 
The IR finiteness still holds, 
since the decomposition changes only an overall finite phase factor of the $S$-matrix element.
However, we cannot move the Kulish-Faddeev dressing operators because of the gauge invariance. 
The KF dressing is a realization of Gauss's law, and we always need a cloud of photons associated with a charged particle. 
Thus, a charged particle without a photon cloud is not allowed.

We have considered uniformly moving charges as \eqref{j_cl} in the classical background model in this paper. 
Although this is enough to look at the leading IR behaviors of QED, 
the true trajectories of particles deviate because of the interaction of them. 
The deviation is proportional to $\log |t|$ in the asymptotic region \cite{Laddha:2018rle, Laddha:2018myi, Sahoo:2018lxl}.
It is discussed in \cite{Laddha:2018myi} that this log behavior is related to the subleading soft factor proportional to $\log \omega$ where $\omega$ is the energy of a soft photon. 
It would be interesting to incorporate the log deviation into the classical current \eqref{j_cl}, and see how the obtained dressed states are related to the subleading soft factor and also sub-subleading factor \cite{Sahoo:2020ryf}. 

In addition, we have assumed that the linear trajectory go through $\vx=0$ at $t=0$. 
We can change it so that $\vx \neq 0$ at $t=0$. 
In \cite{Hamada:2018cjj}, it is shown for the classical case that such a change affects the subleading behavior of the Li\'enard-Wiechert potential and the subleading soft factor in Low's subleading soft theorem \cite{Low:1954kd, Low:1958sn, GellMann:1954kc, Burnett:1967km} is obtained.\footnote{Low's subleading soft factor is $\mathcal{O}(\omega^0)$ and different from the subleading $\log \omega$ term discussed above.}
Thus, if we consider the point-particle current of the shifted trajectories, the dressed states probably include the subleading soft factor. 
Low's subleading soft factor contains the total angular momentum of the charged particles. 
On the other hand, the analysis in \cite{Hamada:2018cjj} is classical, and reproduces only the orbital angular momentum. 
It is unclear whether we can obtain the spin angular momentum in our background current model.  
Dressed states containing the information of this subleading factor is considered in \cite{Choi:2019rlz}.\footnote{Dressed states containing the subleading information are also investigated recently for general Non-abelian gauge theories by using the worldline formalism \cite{Bonocore:2020xuj}.} 
It is worth investigating the dressed states from the shifted trajectories mentioned above. 

It is also important to extend our dressed state formalism to non-abelian gauge theories and gravity.

\section*{Acknowledgement}
The work of H.H. was supported in part by JSPS KAKENHI Grant Number 19J10588.
\appendix
\section{Concrete expression of the phase factor \eqref{eq:defphase} }\label{app:phase}
The phase given in \eqref{eq:defphase} can be written as
\begin{align}
\Phi(t_f,t_i)
=\frac{i}{2}\sum_{n,m} e_{n}e_{m}  (v_{n}\cdot v_{m})\int^{t_f}_{t_i}dt_1 \int^{t_1}_{t_i} dt_2\intkw
    \, \Theta(\eta_{n}t_1)\Theta(\eta_{m}t_2)\left(e^{ik\cdot (v_{n}t_1-v_{m}t_2)}-(c.c)\right),
\end{align}
where $v^{\mu}_{n}:=\frac{p^{\mu}_{n}}{E_{n}}$ is the relativistic velocity for the $n$-th particle, and $\eta_{m}$ takes $-1$ for $m\in I$ and $+1$ for $m\in F$.
After some straightforward calculations,  we obtain the concrete expression of the phase as 
\begin{align}
    \Phi(t_f,t_i)=\intkw \Phi(k;t_f,t_i),
    \label{eq:phase-int}
\end{align}
where
\begin{align}
&\Phi(k;t_f,t_i)
\nn
&=
\sum_{n,m\in I}\Bigg{[}
\frac{e_{n}e_{m}  (v_{n}\cdot v_{m})}{k \cdot v_{m}\ k \cdot \pa{ v_{n}- v_{m}}}\sin\pa{k \cdot \pa{ v_{n}- v_{m}}t_i}
 -\frac{e_{n}e_{m}  (v_{n}\cdot v_{m})}{k \cdot v_{m}\ k \cdot  v_{n}}
\sin\pa{k \cdot  v_{m}t_i}
\Bigg{]}
\nn
&\quad 
+\sum_{n\in F,m\in I}\frac{e_{n}e_{m}  (v_{n}\cdot v_{m})}{k \cdot v_{m}\ k \cdot  v_{n}}
\Big{[}\sin\pa{k \cdot \pa{ v_{n}t_f- v_{m}t_i}}-\sin\pa{k \cdot  v_{n}t_f}+\sin\pa{k \cdot  v_{m}t_i}\Big{]}\nn
&\quad
+\sum_{n,m\in F} \Bigg{[}
-\frac{e_{n}e_{m}  (v_{n}\cdot v_{m})}{k \cdot v_{m}\ k \cdot \pa{ v_{n}- v_{m}}}\sin\pa{k \cdot \pa{ v_{n}- v_{m}}t_f}
 +\frac{e_{n}e_{m}  (v_{n}\cdot v_{m})}{k \cdot v_{m}\ k \cdot  v_{n}}
\sin\pa{k \cdot  v_{m}t_f}
\Bigg{]}.
\label{eq:FKphase}
\end{align}
Part of the phase $\Phi$ diverges in the limit $t_i\rightarrow -\infty, t_f\rightarrow \infty$. More concretely, the first and fifth terms in \eqref{eq:FKphase} with $n=m$ diverge.
The divergent part can be evaluated by decomposing $\Phi(k)$ in \eqref{eq:FKphase} as $\Phi(k)=\Phi_{div}(k)+\Phi_{other}(k)$ where
\begin{align}\label{eq:divphase}
    \Phi_{div}(k;t_f,t_i)
    =-\sum_{n \in  I}\frac{e_{n}^2 m_n^2 }{(k \cdot p_n) E_n}t_i
    +\sum_{n \in  F}\frac{e_{n}^2 m_n^2 }{(k \cdot p_n) E_n}t_f,
\end{align}
and 
\begin{align}
&\Phi_{other}(k;t_f,t_i)
\nn
&=\sum_{\substack{n,m \in I\\  n\neq m}}
\frac{ e_{n}e_{m}  (v_{n}\cdot v_{m})}{k \cdot v_{m}\ k \cdot \pa{ v_{n}- v_{m}}}\sin\pa{k \cdot \pa{ v_{n}- v_{m}}t_i}
 -\sum_{n, m\in I}\frac{e_{n}e_{m}  (v_{n}\cdot v_{m})}{k \cdot v_{m}\ k \cdot  v_{n}}
\sin\pa{k \cdot  v_{m}t_i}
\nn
&+\sum_{n\in F,m\in I}\frac{e_{n}e_{m}  (v_{n}\cdot v_{m})}{k \cdot v_{m}\ k \cdot  v_{n}}
\left[\,\sin\pa{k \cdot \pa{ v_{n}t_f- v_{m}t_i}}-\sin\pa{k \cdot  v_{n}t_f}+\sin\pa{k \cdot  v_{m}t_i}\,\right]\nn
&-\sum_{\substack{n,m \in F\\  n\neq m}}
\frac{e_{n}e_{m}  (v_{n}\cdot v_{m})}{k \cdot v_{m}\ k \cdot \pa{ v_{n}- v_{m}}}\sin\pa{k \cdot \pa{ v_{n}- v_{m}}t_f}
 +\sum_{n,m\in F} \frac{e_{n}e_{m}  (v_{n}\cdot v_{m})}{k \cdot v_{m}\ k \cdot  v_{n}}
\sin\pa{k \cdot  v_{m}t_f}.
\end{align}
Then, the integral 
\begin{align}
    \intkw \Phi_{div}(k;t_f,t_i)
\end{align}
is IR finite for finite $t_i,t_f$,\footnote{This phase is UV divergent, and we need a UV regularization. We do not care about UV divergences in this paper.} but it diverges in the limit $t_i\to -\infty, t_f \to \infty$. 
Thus, the phase is an IR divergent quantity.
The other part 
$
    \intkw \Phi_{other}(k;t_f,t_i)
$
is finite and independent of $t_i,t_f$ because the integral with respect to $\omega$ is given by the following finite integral  
\begin{align}
    \int^\infty_0 \frac{d \omega}{\omega} \sin (\omega a)=\frac{\pi}{2}
\end{align}
which is independent of a constant $a$ appeared in the integrand. 

The divergence does not so matter because $e^{i\Phi}$ is a  phase factor.
The divergent phase takes the form $e^{i g (t_f-t_i)}$ where $g$ is a constant independent of photon states.  
We may absorb the phase by simultaneously redefining final and initial states as $\ket{\alpha(t_f)}_s \to e^{i g t_f}\ket{\alpha(t_f)}_s$ and $\ket{\beta(t_i)}_s \to e^{i g t_i}\ket{\beta(t_i)}_s$. 
Even for the full QED case, the divergent phase would not be important for the following reason.
First the divergent phase can be decomposed as
\begin{align}
e^{i\Phi_{div}(t_f,t_i)}=\prod_{n\in F}e^{i\Phi_n(t_f)}\cdot \prod_{n\in I}e^{-i\Phi_n(t_i)}
\end{align}
where $\Phi_n(t):=\intkw \frac{e_{n}^2 m_n^2 }{(k \cdot p_n) E_n}t$.
Because the phase is diagonal in the momentum space of charged particles, we can absorb the phase by simultaneously redefining the basis for charged sector as $e^{-i\Phi_n(t_i)}\ket{p_n(t_i)}_s \to \ket{p_n(t_i)}_s$, $e^{-i\Phi_n(t_f )}\ket{p_n(t_f)}_s \to \ket{p_n(t_f)}_s$. 
Thus this phase is not relevant for quantum interference.

\section{BRST formalism in the background current model}\label{app:BRST}
We consider the BRST formalism in the background current model where the Lagrangian is given by
\begin{align}
    \mathcal{L}
    =-\frac{1}{4}F_{\mu\nu}F^{\mu\nu}
+j^\mu_{\mathrm{pp}}A_\mu
-\frac{1}{2}\bigl(\partial_{\mu}A^{\mu}\bigr)^{2}+i\,\partial^{\mu}\bar{c}\,\partial_{\mu}c,
\end{align}
where the Nakanishi-Lautrup $B$ field is already integrated out. 
The model has the following BRST symmetry:
\begin{align}
\label{BRST-tr}
    \delta A_\mu= \partial_\mu c, \quad 
    \delta c=0, \quad
    \delta \bar{c}=i\partial_{\mu}A^{\mu}.
\end{align}
We represent the conjugate momentum fields of $A^\mu, c, \bar{c}$ by $\Pi_\mu,  \pc, \bpc$ which are defined as 
\begin{align}
\Pi_0=-\partial_\mu A^\mu, \quad
\Pi_i=F_{0i}, \quad
\pc=-i \partial_0 \bar{c}, \quad 
\bpc=i \partial_0 c.
\end{align}
If we quantize them, 
the canonical commutation relations are 
\begin{align}
[A_\mu^\mathrm{s}(\vx), \Pi_\nu^\mathrm{s}(\vy)] &= i \eta_{\mu\nu} \delta^3(\vx-\vy), \quad \{c^\mathrm{s}(\vx), \pc^\mathrm{s}(\vy)\}=\{\bar{c}^\mathrm{s}(\vx), \bpc^\mathrm{s}(\vy)\}=i \delta^3(\vx-\vy).
\end{align}

The Noether charge generating the BRST transformation \eqref{BRST-tr} is given by
\begin{align}
Q_{BRST}^\mathrm{s}(t)=-\intx \left[i \bpc^\mathrm{s} \Pi^{0\mathrm{s}}-\partial_i c^\mathrm{s} \Pi^{i\mathrm{s}}+ c^\mathrm{s} j_{\mathrm{pp}}^{0}(t,\vx)
\right].
\end{align}
Note that this BRST charge has an explicit time-dependence through the classical background current even in the Schr\"{o}dinger picture.
This BRST charge acts on the fields as
\begin{align}
&[Q_{BRST}^\mathrm{s}, A_{0}^\mathrm{s}]=-\bpc^\mathrm{s},\quad 
[Q_{BRST}^\mathrm{s}, A_{i}^\mathrm{s}(\vx)]=-i\partial_i c^\mathrm{s}, \quad
[Q_{BRST}^\mathrm{s}, \Pi_{\mu}^\mathrm{s}]=0,
\\
&\{Q_{BRST}^\mathrm{s}, \bar{c}^\mathrm{s}\}= \Pi^{0s},\quad
\{Q_{BRST}^\mathrm{s}, \pc^\mathrm{s}\}=-i(\partial_i \Pi^{is}+j_{cl}^{0}),\quad 
\{Q_{BRST}^\mathrm{s}, c^\mathrm{s}\}=
\{Q_{BRST}^\mathrm{s}, \bpc^\mathrm{s}\}=0.
\end{align}
We introduce the annihilation and creation operators as \eqref{a-A}, \eqref{a-P0}, \eqref{a-Pi}, and 
\begin{align}
c^\mathrm{s}(\vx)&= \intkw \left[c(\vk) e^{-i\omega t_s +i \kx} +c^\dagger(\vk) e^{i\omega t_s -i \kx}
\right], 
\\
\bar{c}^\mathrm{s}(\vx)&= \intkw \left[\bar{c}(\vk) e^{-i\omega t_s +i \kx} +\bar{c}^\dagger(\vk) e^{i\omega t_s -i \kx}
\right], 
\\
\pc^\mathrm{s}(\vx)&= -\intk \frac{1}{2}\left[\bar{c}(\vk) e^{-i\omega t_s +i \kx} -\bar{c}^\dagger(\vk) e^{i\omega t_s -i \kx}
\right], 
\\
\bpc^\mathrm{s}(\vx)&= \intk \frac{1}{2}\left[c(\vk) e^{-i\omega t_s +i \kx} -c^\dagger(\vk) e^{i\omega t_s -i \kx}
\right],
\end{align}
where the anti-commutation relations are 
\begin{align}
\{c(\vk), \bar{c}^\dagger(\vkp)\} =i (2\omega)(2\pi)^3  \delta^3(\vk-\vkp),\quad 
\{\bar{c}(\vk), c^\dagger(\vkp)\} =-i(2\omega)(2\pi)^3  \delta^3(\vk-\vkp).
\end{align}
Then, the BRST charge can be written as
\begin{align}
Q_{BRST}^\mathrm{s}(t)=-\intkw \left[
c(\vk) \{k^\mu a_\mu^{\dagger}(\vk) +e^{-i\omega t_s}j^{0}_{\mathrm{pp}}(t,-\vk)\}
+c^\dagger(\vk)  \{k^\mu a_\mu(\vk) +e^{i\omega t_s}j^{0}_{\mathrm{pp}}(t,\vk)\}
\right].
\end{align}

Unlike the full QED, this BRST charge does not commute with the Hamiltonian which is given by
\begin{align}
    H^\mathrm{s}_{tot}(t)=H_0+V^\mathrm{s}(t)+H_{ghost}
\end{align}
where $H_0, V^\mathrm{s}(t)$ are given in \eqref{H0V}, and  the ghost Hamiltonian is 
\begin{align}
    H_{ghost}=\frac{i}{2}\intk \left[c^{\dagger}(\vk) \bar{c}(\vk)-\bar{c}^{\dagger}(\vk) c(\vk)\right].
\end{align}
We actually have
\begin{align}
 [Q^\mathrm{s}_{BRST}(t) , H_{tot}^\mathrm{s}(t)]  
&=i
\intkw \left[
c(\vk)\partial_t j^{0}_{\mathrm{pp}}(t,-\vk)
e^{-i\omega t_s} 
+(c.c.)
\right]
=-i \partial_t Q^\mathrm{s}_{BRST}(t),
\label{noncomu}
\end{align}
where we have used the current conservation $\partial_\mu j^\mu_{\mathrm{pp}}(x)=0$ which means $k_i j^i_{\mathrm{pp}}(t,\vk)=i \partial_t j^0_{\mathrm{pp}}(t,\vk)$.
This non-commutativity \eqref{noncomu} is due to the time-dependence of the background current $j^\mu_{\mathrm{pp}}$.  
Nevertheless, the physical state condition $Q^\mathrm{s}_{BRST}(t)\ket{\psi(t)}_s=0$ is preserved under the time-evolution. 
Indeed, if we have  $Q^\mathrm{s}_{BRST}(t)\ket{\psi(t)}_s =0$, the infinitesimally time-evolved state at $t+\delta t$ satisfies the physical state condition at that time as 
\begin{align}
    Q^{\mathrm{s}}_{BRST}(t+\delta t)U(t+\delta t,t)\ket{\psi(t)}_s 
    &\simeq \delta t \left(\partial_t Q^{\mathrm{s}}_{BRST}(t)-i Q^{\mathrm{s}}_{BRST}(t) H^\mathrm{s}_{tot}(t)
    \right)\ket{\psi(t)}_s
    \nn
   &=-i\delta t  H^\mathrm{s}_{tot}(t) Q^{\mathrm{s}}_{BRST}(t)
    \ket{\psi(t)}_s =0.
\end{align}

\section{Large gauge parameters and the asymptotic charges}
\label{app:as}
We use arbitrary coordinates $\Omega^A (A=1,2)$ to parameterize the celestial two-sphere where the metric components are represented by $\gamma_{AB}$.  
The Minkowski metric is then given by $ds^2=-dt^2+dr^2+r^2\gamma_{AB}d\Omega^Ad\Omega^B$. 

Using the coordinates, the large gauge parameter in the Lorenz gauge is  given by (see \cite{Campiglia:2015qka, Hirai:2018ijc})
\begin{align}
&\epsilon(x)=\int d^2 \Omega' \sqrt{\gamma(\Omega')}\,  G(x;\Omega') \epsilon^{(0)}(\Omega')\,,\label{LGP}\\
& G(x;\Omega') = -\frac{1}{4\pi} \frac{x^\mu x_\mu}{(-t+\hat{x}(\Omega')\cdot \vx )^2} 
\label{bulk_bdry}, 
\end{align}   
where $\hat{x}(\Omega')$ is a three-dimensional unit vector representing a point of the celestial sphere.
This $\epsilon(t,r,\Omega)$ becomes  $\epsilon^{(0)}(\Omega)$ in the limit $t\to +\infty$ with $u=t-r$ fixed and similarly $\epsilon^{(0)}(\bar{\Omega})$ 
in the limit $t\to -\infty$ with $v=t+r$ fixed, where $\bar{\Omega}$ represents the antipodal point of $\Omega$ on the two-sphere, \textit{i.e.}, $\hat{x}(\bar{\Omega})=-\hat{x}(\Omega)$. 

The asymptotic charge in the interaction picture is given by
\begin{align}
\label{eq:app-Q^I_as}
Q_{as}^I[\epsilon]=\intx \left[
-\Pi^{0I} \partial_0 \epsilon -\Pi^{iI}\partial_i \epsilon
+j^0_{\mathrm{pp}}\epsilon
\right]
\end{align}
in the background current model. We now consider the asymptotic limits of this charge.

Using the above expression of $\epsilon$, we can easily find  that 
\begin{align}
    \intx j^0_{\mathrm{pp}}\epsilon=
    \int d^2 \Omega  \sqrt{\gamma(\Omega)}
    \epsilon^{(0)}(\Omega)
    \left[
    \Theta(-t)\sum_{n \in I}\frac{e_n m_n^2}{4\pi(-E_n+\vp_n \cdot \hat{x}(\Omega))^2}
    + \Theta(t)\sum_{n \in F}\frac{e_{n} m_{n}^2}{4\pi(-E_{n}+\vp_{n} \cdot \hat{x}(\Omega))^2}
    \right]
\end{align}
for the classical current \eqref{j_cl}.
We define past and future ``hard charges'' as
\begin{align}
    Q_{hard}^{in}&:=\sum_{n \in I} 
     \int d^2 \Omega  \sqrt{\gamma}
    \epsilon^{(0)}(\Omega)
    \frac{e_n m_n^2}{4\pi(-E_n+\vp_n \cdot \hat{x}(\Omega))^2}, \\
    Q_{hard}^{out}&:=\sum_{n \in F} 
     \int d^2 \Omega  \sqrt{\gamma}
    \epsilon^{(0)}(\Omega)
   \frac{e_{n} m_{n}^2}{4\pi(-E_{n}+\vp_{n} \cdot \hat{x}(\Omega))^2},
\end{align}
and then obtain
\begin{align}
    \intx j^0_{\mathrm{pp}}\epsilon=
    \Theta(-t)Q_{hard}^{in}
    + \Theta(t)Q_{hard}^{out}.
\end{align}
In the asymptotic future region $\mathscr{I}^+$, we have (see, \textit{e.g.}, \cite{Strominger:2017zoo}) 
\begin{align}
    \lim_{t\to +\infty}A_\mu^I(t,r=t-u,\Omega)
    &=-\frac{i}{8\pi^2 t}\int^\infty_0 d\omega\left[a_\mu(\omega \hat{x}(\Omega))e^{-i\omega u}- a_\mu^\dagger(\omega \hat{x}(\Omega))e^{i\omega u}\right].
\end{align}
From this, we can obtain (see, \textit{e.g.}, \cite{Hirai:2018ijc})  \begin{align}
    \lim_{t\to \infty}\intx 
 \Pi^{iI}(t,\vx)\partial_i \epsilon(t,\vx)&= -\frac{1}{8\pi}\lim_{\omega \to 0}\omega
  \int d^2\Omega \sqrt{\gamma}\gamma^{AB}\partial_A  \hat{x}^i\partial_B \epsilon^{(0)}
   \left[ a_i(\omega\hat{x}) +  a_i^{\dagger}(\omega\hat{x})\right]
    \nn
   &:= -Q_{soft}
    .
\end{align}
Similarly, in the past limit, we have
\begin{align}
    \lim_{t\to -\infty}\intx 
 \Pi^{iI}(t,\vx)\partial_i \epsilon(t,\vx)=-Q_{soft}.
\end{align}
Since $\Pi^0$ is the BRST exact, we can ignore the first term in \eqref{eq:app-Q^I_as} on the physical Hilbert space. 
Therefore, we have
\begin{align}
   Q_{as}^{+I}[\epsilon^{(0)}]
:= \lim_{t\to \infty}Q_{as}^I[\epsilon]=Q_{soft} +Q^{out}_{hard},
\end{align}
and
\begin{align}
   Q_{as}^{-I}[\epsilon^{(0)}]
:= \lim_{t\to -\infty}Q_{as}^I[\epsilon]=
    Q_{soft}+Q^{in}_{hard}.
\end{align}
Note that the asymptotic charge becomes the same operator $Q_{soft}$ in the asymptotic future and past  except for the classical hard charge parts.

\section{Fall-off of the Coulomb fields  \label{app:falloff}}
In this appendix, we show that
the integral \eqref{eq:chargeofLW},
\begin{align}\label{eq:chargeofLW2}
    [Q_{as}^I[\epsilon(t_i)],R_{in}(t_i)]
    =\sum_{n\in I}\int\!\! d^3 x F_{0i}^{LW}(t_i,\vx;p_n)\partial^i \epsilon(t_i,\vx)
\end{align}
vanishes in the limit  $t_i\rightarrow -\infty$.
Because the asymptotic region at $t_i\rightarrow -\infty$ consists of the past timelike infinity $i^{-}$ and the past null infinity $\mathscr{I}^{-}$, \eqref{eq:chargeofLW2} can be decomposed as
\begin{align}
 \lim_{t_i\rightarrow-\infty} [Q_{as}^I[\epsilon(t_i)],R_{in}(t_i)] 
 =Q^{LW}_{i^{-}}+Q^{LW}_{\mathscr{I}^{-}}
\end{align}
where
\begin{align}
   &Q^{LW}_{i^{-}}:= \sum_{n\in I}\int_{i^{-}}\!\! d\Sigma^{\mu}_{i^{-}} F_{\mu \nu}^{LW}\, \partial^\nu\epsilon\,,\\
   &Q^{LW}_{\mathscr{I}^{-}}:= \sum_{n\in I}\int_{\mathscr{I}^{-}}\!\! d\Sigma^{\mu}_{\mathscr{I}^{-}} F_{\mu \nu}^{LW}\, \partial^\nu\epsilon.
\end{align}
Above, $d\Sigma^{\mu}_{i^{-}}$ and $d\Sigma^{\mu}_{\mathscr{I}^{-}}$ denote the directed surface elements on $i^{-}$ and $\mathscr{I}^{-}$, respectively.
$Q^{LW}_{i^{-}}$ and $Q^{LW}_{\mathscr{I}^{-}}$ represent the contributions from the Li\'enard-Wiechert potential  to the hard charge and the soft charge, respectively.
First, we can easily show that $Q^{LW}_{\mathscr{I}^{-}}$ vanishes since  Coulomb fields do not reach the null infinities.\footnote{More concretely, $F_{vt}\partial_{v}\epsilon=\mathcal{O}(r^3)$ as $t\rightarrow -\infty$ with $v=t+r$ fixed.}
Next, we evaluate $Q^{LW}_{i^{-}}$.
To focus on the physics at timelike infinity $i^{\pm}$, we introduce the coordinates $(\tau, \rho, \Omega^A)$ on 4d Minkowski spacetime as
\begin{align}
\label{taurhocoord}
ds^{2}=-d\tau^{2}+\tau^{2}\Big{[}\frac{d\rho^{2}}{1+\rho^{2}}+\rho^{2}\gamma_{AB}d\Omega^{A}d\Omega^{B}\Big{]} \ ,
\end{align}
where $\gamma_{AB}$ is a metric on $S^2$.
These coordinates can be obtained by the following coordinate transformation from the Minkowski coordinates,
\begin{align}
    \label{tau:eq}
\tau^{2}=t^{2}-r^{2}\ ,\ \rho=\frac{r}{\sqrt{t^{2}-r^{2}}}\,.
\end{align}
These coordinates are useful to study the asymptotic behaviors of fields at timelike infinity $i^{\pm}$ because $\tau=\pm\infty$ surface spanned by $(\rho, \Omega^{A})$ corresponds to $i^{\pm}$ (e.g., see \cite{Campiglia:2015qka, Strominger:2017zoo}).
The surface element on a $\tau$=\,constant hypersurface is given by
$
d\Sigma_{i^-}^{\mu}
= \delta^{\mu}_{\tau}\,d\rho d^{2}\Omega \frac{|\tau|^3\rho^2\sqrt{\gamma}}{\sqrt{1+\rho^2}}.
$
In these coordinates, the integral \eqref{eq:chargeofLW} can be expressed as
\begin{align}
Q^{LW}_{i^-}
&=\lim_{\tau\rightarrow-\infty}\sum_{n\in I}\int d\rho d^2\Omega\sqrt{\gamma} |\tau| \left(\rho^{2} \sqrt{1+\rho^{2}} F_{\tau \rho}^{LW} \partial_{\rho} \epsilon+\frac{\gamma^{AB}}{\sqrt{1+ \rho^{2}}} F^{LW}_{\tau A} \partial_{B} \epsilon\right)\,.
\label{eq:CLW}
\end{align}
We can also show that the classical Coulomb field falls off around the  timelike infinity as
\begin{align}
 F_{\tau \rho}^{LW}(\tau,\rho,\Omega;p_n)=\mathcal{O}(\tau^{-2})\ ,\    F^{LW}_{\tau A}(\tau,\rho,\Omega;p_n)=\mathcal{O}(\tau^{-2})\,.
\end{align}
Therefore, \eqref{eq:CLW} vanishes.

\section{Soft theorem for charged states}
\label{app:soft}
We will show that the current operator $j^\mu(t,\vk)$ can be approximated by the classical current operator $j^\mu_{\mathrm{pp}}(t,\vk)$ in the soft limit $\vk \to 0$ when it acts on on-shell charged states.

For example, for a 1-particle fermion state $\ket{p,s}=b_s^{\dagger}(\vec{p})\ket{0}$ with charge $e$ , 
the current operator $j_\mu$ acts on this state as 
\begin{align}
    j_\mu(t,\vk)\ket{p,s}=
    \frac{ie}{2E_{p'}}\sum_{s'}\bar{u}^{s'}(p')\gamma_\mu u^\mathrm{s}(p)e^{i(E_{p'}-E_p)t}\ket{p',s'},
\end{align}
where $\vec{p'}=\vec{p}-\vec{k}$.
In the soft limit, we have
$\lim_{\vk\to 0} \bar{u}^{s'}(p')\gamma_\mu u^\mathrm{s}(p)=-2ip_{\mu}\delta_{s',s}$. 
We thus obtain the approximation 
\begin{align}
\label{eq:act_j_ps}
  j_\mu(t,\vk)\ket{p,s} \sim e\frac{p_\mu}{E_p}e^{-i\frac{\vec{p}\cdot \vk}{E_p}t}  \ket{p,s} \qquad \text{for}\quad \vec{k}\sim 0, 
\end{align}
where we have kept the phase factor $e^{-i\frac{\vec{p}\cdot \vk}{E_p}t}$ because we cannot suppose that $\frac{\vec{p}\cdot \vk}{E_p}t$ is small for large $t$.
The right-hand side of \eqref{eq:act_j_ps} is exactly the same as the action of (the momentum representation of) the classical current operator defined in \eqref{eq:j_cl_op} as 
\begin{align}
 e\frac{p^\mu}{E_p}e^{-i\frac{\vec{p}\cdot \vk}{E_p}t}  \ket{p,s}=j^\mu_{\mathrm{pp}} (t,\vk)\ket{p,s}.
\end{align}
Thus, the approximation
\begin{align}
   j^\mu(t,\vk)\ket{p,s} \sim j_{\mathrm{pp}}^\mu (t,\vk)\ket{p,s}
\end{align}
holds for the soft momentum $\vec{k}\sim 0$. The extension to multi-particle states is trivial. Hence, for any free multi-particle states $\ket{\psi}$, we obtain
\begin{align}
\label{eq:state soft thrm}
     j^\mu(t,\vk)\ket{\psi} \sim j_{\mathrm{pp}}^\mu (t,\vk)\ket{\psi}.
\end{align}
This is the soft theorem  at the state level.
The conventional soft photon theorem is easily derived from eq.~\eqref{eq:state soft thrm}.

\bibliographystyle{utphys}
\bibliography{ref_dress}

\providecommand{\href}[2]{#2}\begingroup\raggedright\begin{thebibliography}{10}

\bibitem{Yennie:1961ad}
D.~R. Yennie, S.~C. Frautschi, and H.~Suura, ``{The infrared divergence
  phenomena and high-energy processes},''
\href{http://dx.doi.org/10.1016/0003-4916(61)90151-8}{{\em Annals Phys.}
  {\bfseries 13} (1961) 379--452}.

\bibitem{Weinberg:1965nx}
S.~Weinberg, ``{Infrared photons and gravitons},''
\href{http://dx.doi.org/10.1103/PhysRev.140.B516}{{\em Phys. Rev.} {\bfseries
  140} (1965) B516--B524}.

\bibitem{Weinberg:1995mt}
S.~Weinberg, {\em {The Quantum theory of fields. Vol. 1: Foundations}}.
\newblock Cambridge University Press,
2005.
\newblock

\bibitem{Carney:2017jut}
D.~Carney, L.~Chaurette, D.~Neuenfeld, and G.~W. Semenoff, ``{Infrared quantum
  information},'' \href{http://dx.doi.org/10.1103/PhysRevLett.119.180502}{{\em
  Phys. Rev. Lett.} {\bfseries 119} no.~18, (2017) 180502},
  \href{http://arxiv.org/abs/1706.03782}{{\ttfamily arXiv:1706.03782
  [hep-th]}}.

\bibitem{Bloch:1937pw}
F.~Bloch and A.~Nordsieck, ``{Note on the Radiation Field of the electron},''
\href{http://dx.doi.org/10.1103/PhysRev.52.54}{{\em Phys. Rev.} {\bfseries 52}
  (1937) 54--59}.

\bibitem{Chung:1965zza}
V.~Chung, ``{Infrared Divergence in Quantum Electrodynamics},''
\href{http://dx.doi.org/10.1103/PhysRev.140.B1110}{{\em Phys. Rev.} {\bfseries
  140} (1965) B1110--B1122}.

\bibitem{Greco:1967zza}
M.~Greco and G.~Rossi, ``{A Note on the Infrared Divergence},''
  \href{http://dx.doi.org/10.1007/BF02820731}{{\em Nuovo Cim.} {\bfseries 50}
  (1967) 168}.

\bibitem{Kibble:1968sfb}
T.~Kibble, ``{Coherent Soft-Photon States and Infrared Divergences. I.
  Classical Currents},'' \href{http://dx.doi.org/10.1063/1.1664582}{{\em J.
  Math. Phys.} {\bfseries 9} no.~2, (1968) 315--324}.

\bibitem{Kibble:1969ip}
T.~W.~B. Kibble, ``{Coherent soft-photon states and infrared divergences. ii.
  mass-shell singularities of green's functions},''
\href{http://dx.doi.org/10.1103/PhysRev.173.1527}{{\em Phys. Rev.} {\bfseries
  173} (1968) 1527--1535}.

\bibitem{Kibble:1969ep}
T.~W.~B. Kibble, ``{Coherent soft-photon states and infrared divergences. iii.
  asymptotic states and reduction formulas},''
\href{http://dx.doi.org/10.1103/PhysRev.174.1882}{{\em Phys. Rev.} {\bfseries
  174} (1968) 1882--1901}.

\bibitem{Kibble:1969kd}
T.~W.~B. Kibble, ``{Coherent soft-photon states and infrared divergences. iv.
  the scattering operator},''
\href{http://dx.doi.org/10.1103/PhysRev.175.1624}{{\em Phys. Rev.} {\bfseries
  175} (1968) 1624--1640}.

\bibitem{Kulish:1970ut}
P.~P. Kulish and L.~D. Faddeev, ``{Asymptotic conditions and infrared
  divergences in quantum electrodynamics},''
  \href{http://dx.doi.org/10.1007/BF01066485}{{\em Theor. Math. Phys.}
  {\bfseries 4} (1970) 745}.
[Teor. Mat. Fiz.4,153(1970)].

\bibitem{Greco:1978te}
M.~Greco, F.~Palumbo, G.~Pancheri-Srivastava, and Y.~Srivastava, ``{Coherent
  State Approach to the Infrared Behavior of Nonabelian Gauge Theories},''
  \href{http://dx.doi.org/10.1016/0370-2693(78)90707-4}{{\em Phys. Lett. B}
  {\bfseries 77} (1978) 282--286}.

\bibitem{Ware:2013zja}
J.~Ware, R.~Saotome, and R.~Akhoury, ``{Construction of an asymptotic S matrix
  for perturbative quantum gravity},''
  \href{http://dx.doi.org/10.1007/JHEP10(2013)159}{{\em JHEP} {\bfseries 10}
  (2013) 159},
\href{http://arxiv.org/abs/1308.6285}{{\ttfamily arXiv:1308.6285 [hep-th]}}.

\bibitem{Hirai:2019gio}
H.~Hirai and S.~Sugishita, ``{Dressed states from gauge invariance},''
  \href{http://dx.doi.org/10.1007/JHEP06(2019)023}{{\em JHEP} {\bfseries 06}
  (2019) 023}, \href{http://arxiv.org/abs/1901.09935}{{\ttfamily
  arXiv:1901.09935 [hep-th]}}.

\bibitem{Mirbabayi:2016axw}
M.~Mirbabayi and M.~Porrati, ``{Dressed Hard States and Black Hole Soft
  Hair},'' \href{http://dx.doi.org/10.1103/PhysRevLett.117.211301}{{\em Phys.
  Rev. Lett.} {\bfseries 117} no.~21, (2016) 211301},
\href{http://arxiv.org/abs/1607.03120}{{\ttfamily arXiv:1607.03120 [hep-th]}}.

\bibitem{Gabai:2016kuf}
B.~Gabai and A.~Sever, ``{Large gauge symmetries and asymptotic states in
  QED},'' \href{http://dx.doi.org/10.1007/JHEP12(2016)095}{{\em JHEP}
  {\bfseries 12} (2016) 095},
\href{http://arxiv.org/abs/1607.08599}{{\ttfamily arXiv:1607.08599 [hep-th]}}.

\bibitem{Kapec:2017tkm}
D.~Kapec, M.~Perry, A.-M. Raclariu, and A.~Strominger, ``{Infrared Divergences
  in QED, Revisited},''
  \href{http://dx.doi.org/10.1103/PhysRevD.96.085002}{{\em Phys. Rev.}
  {\bfseries D96} no.~8, (2017) 085002},
\href{http://arxiv.org/abs/1705.04311}{{\ttfamily arXiv:1705.04311 [hep-th]}}.

\bibitem{Choi:2017bna}
S.~Choi, U.~Kol, and R.~Akhoury, ``{Asymptotic Dynamics in Perturbative Quantum
  Gravity and BMS Supertranslations},''
  \href{http://dx.doi.org/10.1007/JHEP01(2018)142}{{\em JHEP} {\bfseries 01}
  (2018) 142},
\href{http://arxiv.org/abs/1708.05717}{{\ttfamily arXiv:1708.05717 [hep-th]}}.

\bibitem{Choi:2017ylo}
S.~Choi and R.~Akhoury, ``{BMS Supertranslation Symmetry Implies Faddeev-Kulish
  Amplitudes},'' \href{http://dx.doi.org/10.1007/JHEP02(2018)171}{{\em JHEP}
  {\bfseries 02} (2018) 171},
\href{http://arxiv.org/abs/1712.04551}{{\ttfamily arXiv:1712.04551 [hep-th]}}.

\bibitem{Carney:2018ygh}
D.~Carney, L.~Chaurette, D.~Neuenfeld, and G.~Semenoff, ``{On the need for soft
  dressing},'' \href{http://dx.doi.org/10.1007/JHEP09(2018)121}{{\em JHEP}
  {\bfseries 09} (2018) 121},
\href{http://arxiv.org/abs/1803.02370}{{\ttfamily arXiv:1803.02370 [hep-th]}}.

\bibitem{Neuenfeld:2018fdw}
D.~Neuenfeld, ``{Infrared-safe scattering without photon vacuum transitions and
  time-dependent decoherence},''
\href{http://arxiv.org/abs/1810.11477}{{\ttfamily arXiv:1810.11477 [hep-th]}}.

\bibitem{Gonzo:2019fai}
R.~Gonzo, T.~Mc~Loughlin, D.~Medrano, and A.~Spiering, ``{Asymptotic Charges
  and Coherent States in QCD},''
  \href{http://arxiv.org/abs/1906.11763}{{\ttfamily arXiv:1906.11763
  [hep-th]}}.

\bibitem{Choi:2019rlz}
S.~Choi and R.~Akhoury, ``{Subleading soft dressings of asymptotic states in
  QED and perturbative quantum gravity},''
  \href{http://dx.doi.org/10.1007/JHEP09(2019)031}{{\em JHEP} {\bfseries 09}
  (2019) 031}, \href{http://arxiv.org/abs/1907.05438}{{\ttfamily
  arXiv:1907.05438 [hep-th]}}.

\bibitem{Choi:2019sjs}
S.~Choi and R.~Akhoury, ``{Magnetic soft charges, dual supertranslations, and
  't Hooft line dressings},''
  \href{http://dx.doi.org/10.1103/PhysRevD.102.025001}{{\em Phys. Rev. D}
  {\bfseries 102} no.~2, (2020) 025001},
  \href{http://arxiv.org/abs/1912.02224}{{\ttfamily arXiv:1912.02224
  [hep-th]}}.

\bibitem{Furugori:2020vdl}
H.~Furugori and S.~Nojiri, ``{Dressed-Asymptotic States From S-matrix and QED
  Large Gauge Symmetry},'' \href{http://arxiv.org/abs/2007.02518}{{\ttfamily
  arXiv:2007.02518 [hep-th]}}.

\bibitem{Dybalski:2017mip}
W.~Dybalski, ``{From Faddeev\textendash{}Kulish to LSZ. Towards a
  non-perturbative description of colliding electrons},''
  \href{http://dx.doi.org/10.1016/j.nuclphysb.2017.10.018}{{\em Nucl. Phys. B}
  {\bfseries 925} (2017) 455--469},
  \href{http://arxiv.org/abs/1706.09057}{{\ttfamily arXiv:1706.09057
  [hep-th]}}.

\bibitem{Kugo:1977zq}
T.~Kugo and I.~Ojima, ``{Manifestly Covariant Canonical Formulation of
  Yang-Mills Field Theories: Physical State Subsidiary Conditions and Physical
  S Matrix Unitarity},''
\href{http://dx.doi.org/10.1016/0370-2693(78)90765-7}{{\em Phys. Lett.}
  {\bfseries 73B} (1978) 459--462}.

\bibitem{Bagan:1999jf}
E.~Bagan, M.~Lavelle, and D.~McMullan, ``{Charges from dressed matter:
  Construction},'' \href{http://dx.doi.org/10.1006/aphy.2000.6048}{{\em Annals
  Phys.} {\bfseries 282} (2000) 471--502},
\href{http://arxiv.org/abs/hep-ph/9909257}{{\ttfamily arXiv:hep-ph/9909257
  [hep-ph]}}.

\bibitem{He:2014cra}
T.~He, P.~Mitra, A.~P. Porfyriadis, and A.~Strominger, ``{New Symmetries of
  Massless QED},'' \href{http://dx.doi.org/10.1007/JHEP10(2014)112}{{\em JHEP}
  {\bfseries 10} (2014) 112},
\href{http://arxiv.org/abs/1407.3789}{{\ttfamily arXiv:1407.3789 [hep-th]}}.

\bibitem{Campiglia:2015qka}
M.~Campiglia and A.~Laddha, ``{Asymptotic symmetries of QED and Weinberg's soft
  photon theorem},'' \href{http://dx.doi.org/10.1007/JHEP07(2015)115}{{\em
  JHEP} {\bfseries 07} (2015) 115},
\href{http://arxiv.org/abs/1505.05346}{{\ttfamily arXiv:1505.05346 [hep-th]}}.

\bibitem{Kapec:2015ena}
D.~Kapec, M.~Pate, and A.~Strominger, ``{New Symmetries of QED},''
  \href{http://dx.doi.org/10.4310/ATMP.2017.v21.n7.a7}{{\em Adv. Theor. Math.
  Phys.} {\bfseries 21} (2017) 1769--1785},
\href{http://arxiv.org/abs/1506.02906}{{\ttfamily arXiv:1506.02906 [hep-th]}}.

\bibitem{Hirai:2018ijc}
H.~Hirai and S.~Sugishita, ``{Conservation Laws from Asymptotic Symmetry and
  Subleading Charges in QED},''
  \href{http://dx.doi.org/10.1007/JHEP07(2018)122}{{\em JHEP} {\bfseries 07}
  (2018) 122},
\href{http://arxiv.org/abs/1805.05651}{{\ttfamily arXiv:1805.05651 [hep-th]}}.

\bibitem{Bieri:2013hqa}
L.~Bieri and D.~Garfinkle, ``{An electromagnetic analogue of gravitational wave
  memory},'' \href{http://dx.doi.org/10.1088/0264-9381/30/19/195009}{{\em
  Class. Quant. Grav.} {\bfseries 30} (2013) 195009},
\href{http://arxiv.org/abs/1307.5098}{{\ttfamily arXiv:1307.5098 [gr-qc]}}.

\bibitem{Susskind:2015hpa}
L.~Susskind, ``{Electromagnetic Memory},''
\href{http://arxiv.org/abs/1507.02584}{{\ttfamily arXiv:1507.02584 [hep-th]}}.

\bibitem{Laddha:2018rle}
A.~Laddha and A.~Sen, ``{Gravity Waves from Soft Theorem in General
  Dimensions},'' \href{http://dx.doi.org/10.1007/JHEP09(2018)105}{{\em JHEP}
  {\bfseries 09} (2018) 105},
\href{http://arxiv.org/abs/1801.07719}{{\ttfamily arXiv:1801.07719 [hep-th]}}.

\bibitem{Laddha:2018myi}
A.~Laddha and A.~Sen, ``{Logarithmic Terms in the Soft Expansion in Four
  Dimensions},'' \href{http://dx.doi.org/10.1007/JHEP10(2018)056}{{\em JHEP}
  {\bfseries 10} (2018) 056},
\href{http://arxiv.org/abs/1804.09193}{{\ttfamily arXiv:1804.09193 [hep-th]}}.

\bibitem{Sahoo:2018lxl}
B.~Sahoo and A.~Sen, ``{Classical and Quantum Results on Logarithmic Terms in
  the Soft Theorem in Four Dimensions},''
  \href{http://dx.doi.org/10.1007/JHEP02(2019)086}{{\em JHEP} {\bfseries 02}
  (2019) 086}, \href{http://arxiv.org/abs/1808.03288}{{\ttfamily
  arXiv:1808.03288 [hep-th]}}.

\bibitem{Sahoo:2020ryf}
B.~Sahoo, ``{Classical Sub-subleading Soft Photon and Soft Graviton Theorems in
  Four Spacetime Dimensions},''
  \href{http://arxiv.org/abs/2008.04376}{{\ttfamily arXiv:2008.04376
  [hep-th]}}.

\bibitem{Hamada:2018cjj}
Y.~Hamada and S.~Sugishita, ``{Notes on the gravitational, electromagnetic and
  axion memory effects},''
  \href{http://dx.doi.org/10.1007/JHEP07(2018)017}{{\em JHEP} {\bfseries 07}
  (2018) 017},
\href{http://arxiv.org/abs/1803.00738}{{\ttfamily arXiv:1803.00738 [hep-th]}}.

\bibitem{Low:1954kd}
F.~E. Low, ``{Scattering of light of very low frequency by systems of spin
  1/2},''
\href{http://dx.doi.org/10.1103/PhysRev.96.1428}{{\em Phys. Rev.} {\bfseries
  96} (1954) 1428--1432}.

\bibitem{Low:1958sn}
F.~E. Low, ``{Bremsstrahlung of very low-energy quanta in elementary particle
  collisions},''
\href{http://dx.doi.org/10.1103/PhysRev.110.974}{{\em Phys. Rev.} {\bfseries
  110} (1958) 974--977}.

\bibitem{GellMann:1954kc}
M.~Gell-Mann and M.~L. Goldberger, ``{Scattering of low-energy photons by
  particles of spin 1/2},''
\href{http://dx.doi.org/10.1103/PhysRev.96.1433}{{\em Phys. Rev.} {\bfseries
  96} (1954) 1433--1438}.

\bibitem{Burnett:1967km}
T.~H. Burnett and N.~M. Kroll, ``{Extension of the low soft photon theorem},''
\href{http://dx.doi.org/10.1103/PhysRevLett.20.86}{{\em Phys. Rev. Lett.}
  {\bfseries 20} (1968) 86}.

\bibitem{Bonocore:2020xuj}
D.~Bonocore, ``{Asymptotic dynamics on the worldline for spinning particles},''
  \href{http://arxiv.org/abs/2009.07863}{{\ttfamily arXiv:2009.07863
  [hep-th]}}.

\bibitem{Strominger:2017zoo}
A.~Strominger, ``{Lectures on the Infrared Structure of Gravity and Gauge
  Theory},''
\href{http://arxiv.org/abs/1703.05448}{{\ttfamily arXiv:1703.05448 [hep-th]}}.

\end{thebibliography}\endgroup
\end{document}